\documentclass[journal,10pt]{IEEEtran}

\usepackage{algorithm, algorithmic}

\ifCLASSOPTIONcompsoc

\usepackage[nocompress]{cite}
\else

\usepackage{cite}
\fi

\usepackage{amsthm}

\ifCLASSINFOpdf
\usepackage[pdftex]{graphicx}
\graphicspath{{../pdf/}{../jpeg/}}
\DeclareGraphicsExtensions{.pdf,.jpeg,.png}
\else
\usepackage[dvips]{graphicx}
\graphicspath{{../eps/}}
\DeclareGraphicsExtensions{.eps}
\fi

\usepackage{bm}
\usepackage{color}
\usepackage{amsmath,amssymb,amsfonts,mathrsfs,amsthm,graphicx,epstopdf,bm}
\usepackage{esint,epsfig, subfigure,soul}
\usepackage[usenames,dvipsnames,svgnames]{xcolor}
\usepackage{cite}
\usepackage{scalerel}
\usepackage[normalem]{ulem}
\allowdisplaybreaks  
\usepackage{cleveref}

\newtheorem{lemma}{Lemma}
\newtheorem{remark}{Remark}
\theoremstyle{remark}

\newtheoremstyle{mystyle}
  {}
  {}
  {\itshape}
  {}
  {\bfseries}
  {.}
  { }
  {}

\theoremstyle{mystyle}
\newtheorem{theorem}{Theorem}

\usepackage{amsthm}
\usepackage{verbatim}
\DeclareMathOperator*{\argmin}{argmin}

\usepackage{mathtools}

\usepackage[utf8]{inputenc}

\begin{document}
    \title{Terahertz Communication Multi-UAV-Assisted Mobile Edge Computing System}
	
	\author{Heekang~Song,~\IEEEmembership{Student~Member,~IEEE},
		$^\dagger$Hyowoon~Seo,~\IEEEmembership{Member,~IEEE},
		and~Wan~Choi,~\IEEEmembership{Fellow,~IEEE}
		\thanks{
			H. Song is with the School of Electrical Engineering, Korea Advanced Institute of Science and Technology (KAIST), Daejeon 34141, Korea  (e-mail: hghsong@kaist.ac.kr). 
		}
		\thanks{
			$^\dagger$H. Seo is with the Department of Electronics and Communications Engineering, Kwangwoon University, Seoul 01897, Korea (e-mail: hyowoonseo@kw.ac.kr).
		}
		\thanks{	
			W.~Choi is with the Department of Electrical and Computer Engineering and the Institute of New Media and Communications, Seoul National University (SNU), Seoul 08826, Korea (e-mail:  wanchoi@snu.ac.kr). (\emph{Corresponding Authors:  Hyowoon Seo and Wan Choi})
		}
	}
	\maketitle
	\thispagestyle{empty}
	\begin{abstract}
        Mobile edge computing (MEC) and terahertz (THz)-enabled unmanned aerial vehicle (UAV) communication systems are gaining significant attention for improving user service delays in future mobile networks. This article introduces a novel multi-UAV-aided MEC system operating at THz frequencies to minimize expected user service delays, including communication and computation latency. We address this challenge by jointly optimizing UAV relay selection, power control, positioning, and user-resource association for task offloading and resource allocation. To tackle the problem's complexities, we decompose it into four subproblems, each solved optimally with our proposed algorithm. An iterative penalty dual decomposition (PDD) algorithm approximates the original problem's solution. Numerical results demonstrate that our PDD-based approach outperforms baseline algorithms in terms of expected user service delay.
		\end{abstract}
	
	\begin{IEEEkeywords}
		Mobile edge computing, terahertz communications, unmanned aerial vehicle, penalty dual decomposition.
	\end{IEEEkeywords}
	
	\IEEEpeerreviewmaketitle
	
	\section{Introduction}
    The rapid advancements in software and hardware technologies have led to a proliferation of internet-connected devices, such as smart cameras, smartphones, and smart vehicles, all relying on services like gaming, video streaming, and smart city integrations. This growth has driven the widespread adoption of the Internet of Things (IoT), significantly increasing data generation and necessitating extensive storage and computational processing. This surge in data is expected to spur innovations in VR, AR, and high-definition video streaming but also strains existing network infrastructures, requiring enhanced communication and computational capabilities. Many IoT devices, including sensors and mobile devices, have limited capacities for communication, computing, and storage, affecting the performance of these demanding applications. Initially, cloud computing was adopted to manage these resource-intensive tasks using centralized cloud servers \cite{ref:Dinh}. However, the increasing physical distance between clients and these servers can degrade the Quality-of-Experience (QoE) for clients, especially in wireless communications. 

    Recently, \emph{mobile edge computing (MEC)} has gained traction as an effective approach to improve QoE by offloading tasks to distributed edge servers, utilizing idle computational resources \cite{ref:Shi, ref:Mao}, though at the expense of communication resources. A major challenge in MEC is minimizing user service delay, which includes both communication and computation delays. Communication delay depends on user-edge server channel conditions, while computation delay relies on the availability of idle computing resources at the edge server. Previous research has focused on task offloading techniques to reduce user service delay through scheduling, server deployment, and more \cite{ref:Liu2, ref:Chen2, ref:Zhang3, ref:Wu, ref:Song}. However, the increasing wireless traffic and limited spectral resources in modern gigahertz (GHz) wireless communication systems hinder meeting QoE requirements. Therefore, terahertz (THz) wireless communication (0.1-10 THz) has emerged as a promising solution for wireless MEC to meet these demands \cite{ref:Du, ref:Liu, ref:Chaccour}.	
	
	\begin{figure}[t]
	\centering
	{\includegraphics[width=0.9\columnwidth]{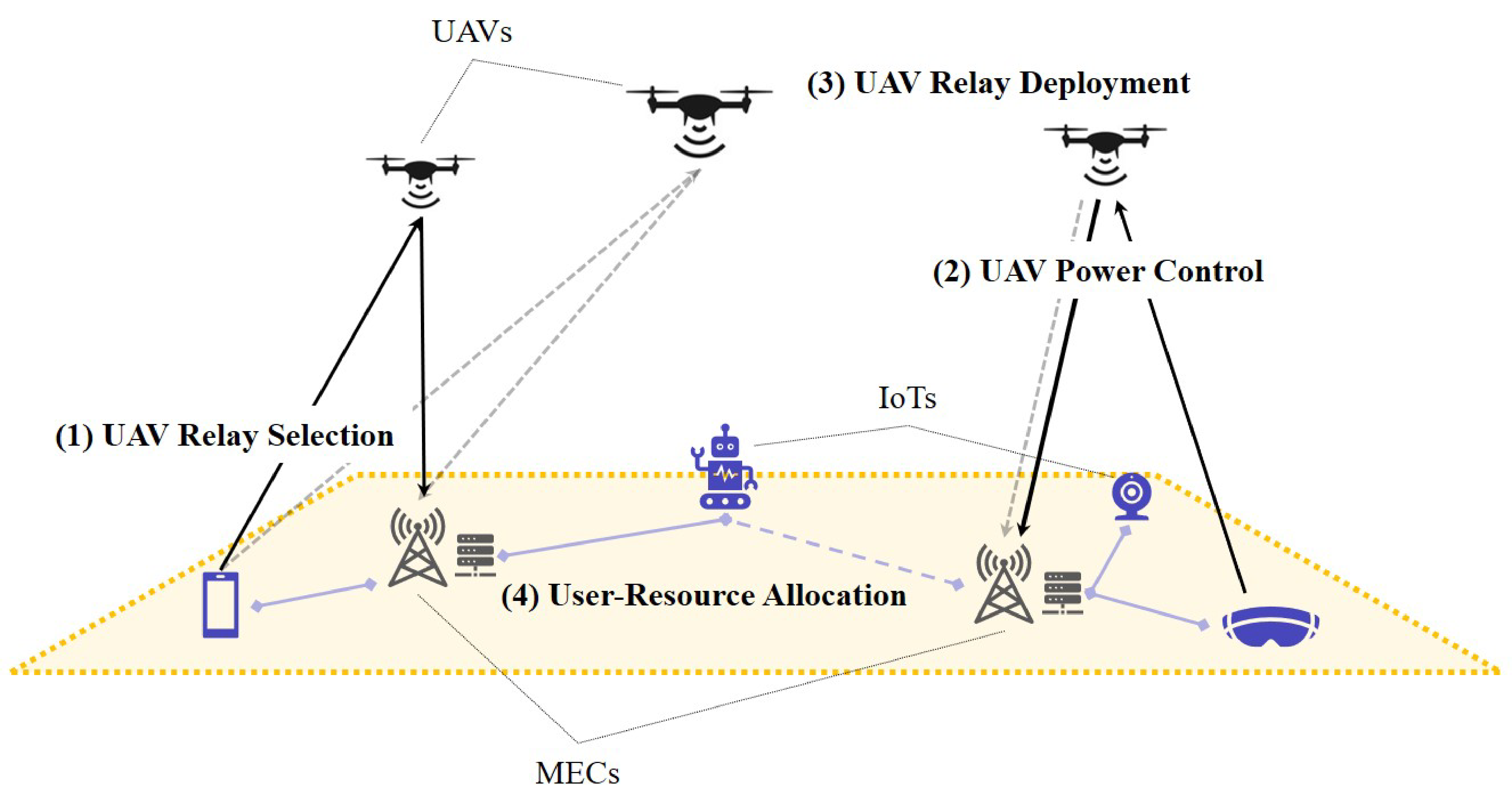}
 } 
	\caption{An illustration of THz-enabled MEC system with multi-UAV communication relays.}
	\label{fig:sys}
	\end{figure}
    THz communication faces challenges such as susceptibility to blockage, high path loss, and molecular absorption loss \cite{ref:Jornet, ref:Ye}. Ensuring stable communication pathways in THz-enabled MEC is crucial, and traditional methods with numerous base stations are inefficient in THz scenarios. Instead, UAVs effectively address communication reliability challenges in THz-enabled MEC due to their flexibility and mobility \cite{ref:Azari}. Various publications discuss THz-UAV network challenges \cite{ref:Azari,  ref:Amodu2, ref:Saeed}. Recent research \cite{ref:Saeed} on aerial communication in the THz band shows promise for integration into vehicular aerial communication systems. Strategic UAV relay placement can mitigate blockages, shorten transmission distances, and enhance communication link quality \cite{ref:Mamaghani, ref:Xu, ref:Pan, ref:Hassan}.
	
    In this context, our primary goal is to design a THz-enabled MEC system with UAV communication relays to meet strict delay requirements for computation-intensive tasks. Key design challenges include optimizing (1) UAV relay selection, (2) UAV transmission power control, (3) UAV relay deployment, and (4) user-resource association methods to meet predefined user service delay requirements. Unlike GHz communications, THz channel path loss involves molecular absorption loss, dependent on link distance \cite{ref:Han}, affecting UAV relay selection, placement, and user offloading decisions. Additionally, communication path loss varies with frequency bands \cite{ref:Shafie}, necessitating optimization of transmit power control and sub-band allocation to mitigate path loss and meet QoE expectations. Optimizing user-resource associations also helps minimize computation delay by balancing the load across MEC servers. Addressing these aspects through the joint design of (1)-(4) is crucial yet challenging for establishing a THz-enabled MEC system with multi-UAV communication relays.

	\subsection{Related Works}

    Reducing user service delays and energy consumption in MEC systems is a complex challenge. Previous studies have shown that this can be achieved through task offloading \cite{ref:Liu2, ref:Chen2, ref:Zhang3, ref:Wu, ref:Song, ref:Fu, ref:Zhan, ref:Yang2}, which introduces transmission and computation delays. Strategies like optimizing server placement to minimize communication distances and scheduling MEC servers to prevent overloading reduce these delays. For instance, \cite{ref:Liu2} proposed a stochastic offloading scheduling rule using a Markov chain model to minimize average task delay for a single user. Extending to multi-user scenarios, \cite{ref:Chen2} used a game-theoretic approach to formulate a multi-user task offloading game to find Nash equilibrium. \cite{ref:Zhang3} presented a mechanism to minimize energy consumption considering both communication and computation energy. However, these techniques were studied in single base station scenarios, limiting their applicability to dense networks with multiple base stations and users. To address this, \cite{ref:Wu} focused on minimizing overall computation delays in scenarios with multiple edge servers by optimizing task offloading and transmission times.

    To further enhance the QoE of computation offloading, Fu et al.\cite{ref:Fu}  explored scenarios where a single terrestrial relay assists IoT devices, effective only with sufficient Line of Sight (LoS), but limited by high installation costs and reduced flexibility. \cite{ref:Yang2} proposed a multi-UAV-aided MEC framework for IoT networks, with UAVs as mobile-edge nodes offering computing services, enhanced by differential evolution-based UAV deployment and DRL-aided task scheduling for load balancing and QoS fulfillment. \cite{ref:Zhan}  introduced a multi-UAV-enabled MEC framework to maximize service coverage for IoT devices under time constraints by optimizing UAV trajectories, service indicators, resource allocation, and computation offloading. Most UAV-aided MEC networks perform computations within UAVs, which is inefficient for applications requiring significant computing power, such as AI. While existing MEC research focuses on minimizing delays for single task requests in computation-abundant scenarios, services like smart farms and smart cities require continuous computation requests with limited computing capacity. In such cases, a computation queue model addressing long-term delays is practical. \cite{ref:Song} recently examined a computation queue model in a computation-limited scenario, proposing a genetic algorithm-based strategy for joint task offloading and server deployment to minimize expected user service delays, demonstrated using real data from Oulu, Finland.    

    Despite efforts to enhance MEC systems, current communication systems may face spectrum limitations. Recent research has focused on leveraging ultra-wideband THz links for MEC-assisted wireless VR scenarios to reduce latency and energy consumption. \cite{ref:Du} optimized rendering offloading and transmit power control to minimize long-term energy consumption. \cite{ref:Liu} proposed long-term QoE optimization for indoor MEC-enabled THz VR networks using learning-based prediction, rendering, and transmission. \cite{ref:Chaccour} explored THz's potential for high-rate, reliable, low-latency wireless VR communication. These studies primarily focused on indoor wireless VR scenarios, limiting the broader application of MEC.

    THz communications suffer from fragility due to blockage, high path loss, and molecular absorption loss, resulting in limited range. To extend THz coverage, studies have explored UAV relay-aided THz communications \cite{ref:Mamaghani, ref:Xu, ref:Pan, ref:Hassan}. For example, \cite{ref:Mamaghani} optimized secrecy energy efficiency in an untrusted UAV-relay system. \cite{ref:Xu} minimized delay through joint optimization of UAV location, bandwidth, and power. \cite{ref:Pan} aimed to maximize sum rates in UAV-aided THz communication. While these studies mainly considered single UAV relays, \cite{ref:Hassan} delved into a multi-UAV communication architecture to maximize overall throughput.
    
    A recent investigation \cite{ref:Azari2} explored a mmWave/THz-enabled cellular MEC system for UAV computation requests, focusing on energy-efficient methodologies for a single UAV. Similarly, \cite{ref:Wang} examined UAV placement, resource allocation, and computation offloading in UAV-assisted MEC systems using a deep reinforcement learning approach, also centered on a single UAV and aimed at minimizing short-term delays for single-task users. To our knowledge, our study is the first to introduce a comprehensive framework for THz-enabled MEC systems with multi-UAV communication relays, aiming to minimize long-term user service delays.
	
	\subsection{Contributions and Organization}
	The major contributions of this paper are listed below.
	\begin{enumerate}
	    \item Leveraging the ultra-wideband THz frequencies and cost-effective UAV communication relays, we present a novel architecture for a THz-enabled MEC system with multi-UAV relays, effectively mitigating IoT devices' limited power and severe coverage limitations.

        \item To minimize user service delay, we jointly optimize UAV relay selection, UAV transmission power control, UAV relay deployment, and user-resource association. Our optimization problem is highly complicated due to the THz link characteristics influenced by communication distance and frequency.

        \item We address the problem's complexity with an iterative penalty dual decomposition (PDD) algorithm, leveraging the mathematical findings of each subproblem convexity. We derive closed-form expressions for two subproblems as the algorithm converges to a stationary point. Additionally, we analyze convergence and computational complexity, ensuring polynomial-time convergence to at least a suboptimal solution.

        \item Numerical results confirm the effectiveness of our design compared to other methods. They emphasize the importance of optimizing both communication and computation delays in a THz-enabled MEC system with multi-UAV communication relays.
	\end{enumerate}

    The rest of this paper is organized as follows: Section \ref{sec:system} describes our THz channel, data transmission, and task computation models. We derive the user service delay and formulate the joint optimization problem to minimize the expected user service delay under delay constraints. In Section \ref{sec:algorithm}, we propose a PDD-based iterative algorithm to solve this problem. Section \ref{sec:simulation} presents the performance evaluation of the proposed algorithm and comparisons with benchmark schemes. Finally, conclusions are drawn in Section \ref{sec:conclusion}.


	\section{System Model} \label{sec:system}

	This section describes the system model under study. As shown in Fig. \ref{fig:sys}, consider a THz communication-based UAV-assisted MEC system, wherein $M$ UAVs are deployed to relay the data transmissions from $I$  Internet of Things (IoTs) to $J$ MEC servers. The finite sets of UAVs, IoTs, and MECs are denoted by $\mathcal{M}\triangleq\{1,...,M\}$, $\mathcal{I}\triangleq\{1,...,I\}$, and $\mathcal{J}\triangleq\{1,...,J\}$, respectively. Each IoT $i \in \mathcal{I}$ generates computation-intensive and delay-sensitive task requests with an arrival rate $\lambda_i$ periodically, which follows the Poisson process \cite{ref:Song, ref:Mao}. The IoTs possess no computational power to handle the tasks by themselves, and thus they offload the tasks to the MEC servers and receive the task results.
	
	The location of a ground entity $k \in \mathcal{I} \cup \mathcal{J}$ is denoted by the 3-dimensional Cartesian coordinates $\mathbf{u}_k = [x_k,y_k, 0]$, and that of UAV $m \in \mathcal{M}$ is denoted by $\mathbf{q}_m = [x_m, y_m, z_m]$, while the altitude of UAVs is fixed at $z_m = H$. 	The nature of THz radiation, such as a high molecular absorption loss and vulnerability to blockage \cite{ref:Han}, hinders direct communication between two entities that are located far apart or using low transmit power. On balance, it is beneficial that the multiple UAVs act as an \emph{aerial relay} node to overcome the obstacles. As a consequence, there exist three different types of directional communications, i.e., IoT-to-MEC, IoT-to-UAV, and UAV-to-MEC, where the set of all possible communication pairs are denoted by $\mathcal{C}=\{(k, l) ~|~ (k,l) \in (\mathcal{I}, \mathcal{J}) \cup (\mathcal{I}, \mathcal{M}) \cup (\mathcal{M}, \mathcal{J})\}$. The Euclidean distance between a communication pair $(k,l)\in \mathcal{C}$ is denoted by $d_{(k,l)} = \sqrt{(x_k - x_l)^2 + (y_k - y_l)^2 + (z_k-z_l)^2}$.

	\subsection{THz Communication Channel Model}
	In general, THz bands are classified by two regions according to the absorption loss, i.e., THz  absorption loss peak regions and ultra-wideband THz transmission windows \cite{ref:Jornet, ref:Shafie}.
	Due to the existence of high signal attenuation at the molecular absorption loss peak region, we only focus on the ultra-wideband THz transmission windows, rather than both, implying that no molecular absorption peak occurs in the considered transmission window. In our scenario, the spectrum of interest is divided into $U$ sub-bands, i.e., $\mathcal{U} \triangleq \{1, ..., U\}$, with equal bandwidth $B$. Thus, the center frequency $f_u$ of a sub-band $u \in \mathcal{U}$ is
	\begin{gather*}
		f_u = f_o + \left(u-\frac{1}{2}\right)B,
	\end{gather*}
	where $f_o$ represents the lowest carrier frequency of the spectrum under consideration. The THz channel propagation is mostly determined by free space spreading and molecular absorption losses. Since the LoS path is dominant \cite{ref:Shafie, ref:Jornet, ref:Han}, it is assumed that the impact of non-LoS is negligible. Hence, the channel gain of a communication pair $c \in \mathcal{C}$ through the frequency $f_u$ is obtained as
	\begin{gather*} 
		\big|h_{c}^{u}\big|^2 = \bigg(\frac{s_\text{light}}{4\pi f_u d_{c}}\bigg)^2 e^{-K(f_u)d_{c}},
	\end{gather*}
	where $s_\text{light}$ is the speed of light, $d_{c}$ is the distance between the communication pair $c$, and $K(f_u)$ is the molecular absorption coefficient at the sub-band $u$. Note that $e^{-K(f_u)d_{c}}$ is the molecular absorption loss, which incorporates the effect of the oxygen and water vapor molecule absorbing the signal energy. The absorption coefficient $K(f_u)$ can be calculated from $K(f_u) = \frac{p}{\overline{p}} \frac{\overline{T}}{T} \sum_{i,g} Q^{i,g} \sigma^{i, g} (f_u)$, wherein $p$ and $T$ are the pressure and the temperature of transmission environment, respectively, while $\overline{p}$ and $\overline{T}$ represent the standard pressure and temperature, respectively. Further, $Q^{i,g}$ and $\sigma^{i, g} (f_u)$ are the total number of molecules per unit volume and the absorption cross section for the isotopologue $i$ of gas $g$ at frequency $f_u$ \cite{ref:Jornet}. Nevertheless, with the aid of HITRAN database \cite{ref:HITRAN}, the absorption coefficient can be readily attained, without any complex calculations
	
    Meanwhile, blockages substantionally impact the performance  of THz communication systems, a consequence inherently linked to the properties of THz radiation. Therefore, a thorough understanding of blockage effects is indispensable for deeper insights into the characteristics of THz communications.
    As described in \cite{ref:Shafie}, the non-blockage probability is modeled using a statistical framework involving randomly moving blockers characterized by height $h_\mathsf{b}$ and radius $\tau_\mathsf{b}$. These blockers are distributed according to a Poisson point process (PPP) with uniform intensity $\beta_{\mathsf{b}}$. The probability of non-blockage for the direct communication pair $c$ becomes 
    $$
    Pr^{\mathsf{nb}} (d_c) = \zeta e^{ - \delta_{\mathsf{b}} d_c}, ~~ \forall c \in (\mathcal{I}, \mathcal{J}),
    $$
    where $\zeta=e^{-2 \beta_{\mathsf{b}} \tau_\mathsf{b}^2}$ and $\delta_{\mathsf{b}} = 2 \beta_{\mathsf{b}} \tau_{\mathsf{b}} (h_\mathsf{b} - h_\mathsf{IoT})/(h_\mathsf{MEC} - h_\mathsf{IoT})$, with $h_\mathsf{MEC}$ and $h_\mathsf{IoT}$ representing the heights of MEC and IoT, respectively. 
    We assume that a user only transmits data through its associated links when the links are unobstructed by dynamic blockers, deemed impenetrable \cite{ref:Shafie, ref:blockage}; concurrently, UAV communication relays maintain robustness and are unaffected by such blockages.
    
     Therefore, the long-term throughput of the communication pair $c \in \mathcal{C}$ when using $u$-th sub-band is
	\begin{gather*} 
		R_{c}^u = \begin{cases}
            Pr^{\mathsf{nb}} (d_c) \cdot B \log_2 \bigg(1+\frac{P_c |h_{c}^u|^2}{B N_0}\bigg), ~~\forall c \in (\mathcal{I}, \mathcal{J}), \\ 
            B \log_2 \bigg(1+\frac{P_c |h_{c}^u|^2}{B N_0}\bigg), ~~~~ \forall c \in  (\mathcal{I}, \mathcal{M}) \cup (\mathcal{M}, \mathcal{J}), 
            \end{cases}
	\end{gather*}
	where $P_c$ is the transmit power of source unit in communication pair $c$ (e.g., the transmit power of IoT or UAV) and $N_0$ is the noise spectral density.

	 \subsection{Data Transmission Model}
	 In the considered network, there exists two different types of communication path that IoTs can use to offload the tasks. One of them is a \emph{direct path}, e.g., a IoT $i \in \mathcal{I}$ transmitting a computing task to an MEC server $j \in \mathcal{J}$ directly. The other is a relay path, e.g., a IoT $i \in \mathcal{I}$ first communicates with a selected UAV $m \in \mathcal{M}$ to route a computing task, then the UAV relays to an MEC server $j \in \mathcal{J}$.
 	For the sake of convenience, indicator variables
 	\begin{gather*}
 		\alpha_{i,m} = \begin{cases}
 			1, & \text{if IoT $i$ selects UAV $m$ as a relay},\\
 			0, & \text{otherwise}.
 		\end{cases}
 	\end{gather*}
 	$\forall i \in \mathcal{I}, \forall m\in \mathcal{M}$, are used hereinafter. Correspondingly, note that $\alpha_i =\sum_{m} \alpha_{i,m}$, $\forall i\in\mathcal{I}$ indicates the type of communication path that the IoTs are using, for example, $\alpha_i =1$ if the IoT $i$ utilizes the UAV relay and $\alpha_i =0$ if the direct path is selected to transmit. Here, it is assumed that the associated pair $(i,j) \in \mathcal{I}\times\mathcal{J}$ for offloading can only get help from at most one relaying UAV, leading to a constraint
 	 \begin{gather} \label{eq:const1}
 	 	\alpha_{i} \leq 1, \quad \forall i \in \mathcal{I}.	
 	 \end{gather}
   Note that the collision avoidance is critical when deploying UAVs. As UAVs typically converge to positions between IoT devices and MEC servers to minimize communication delays, we implement a specific constraint \eqref{eq:const1} to prevent UAVs from being placed at the same location unless the user location and associated MEC are identical. Additionally, our pre-planning strategy ensures that, even in cases of overlapping placements, UAVs are positioned slightly apart to prevent collisions.
 	
   Each UAV splits and adaptively allocates the transmit power for the IoTs that are using it as a relay. Let $P_{i,m}$ be the transmit power allocated by   UAV $m$ that is assigned to IoT $i$, accordingly.
 	 \begin{gather} \label{eq:const12}
 	 	\quad \sum_{i} P_{i, m} \leq P_\mathsf{UAV}, \quad \forall m \in \mathcal{M}.	
 	 \end{gather}
 	 where $P_\mathsf{UAV}$ is the total transmit power of each UAV.
 	 	
 	In the meantime, each IoT chooses an MEC server to which it will offload the tasks and sub-band that it will be used for communication. To avoid any possible interference issues, we state in advance that the orthogonal multiple access (OMA) is considered \cite{ref:Song} throughout this work.
    \footnote{
    Utilizing frequency reuse techniques boosts spectral efficiency in this scenario. However, the current limitations of THz band digital processors and their signal processing overhead hinder these benefits \cite{ref:Han}. Fortunately, the ultra-wideband transmission window in THz communications can achieve high throughput without frequency reuse techniques, thanks to the availability of hundreds of GHz \cite{ref:Shafie}. 
    } 
    In other words, no same sub-band is assigned to more than one IoT. Define by
 	\begin{gather*}
 		z_{j,i}^u = \begin{cases}
 			1, & \text{if IoT $i$ offloads to MEC $j$ via sub-band $u$},\\
 			0, & \text{otherwise},
 		\end{cases}
 	\end{gather*}
 	the indicator of association between the IoT $i$ and MEC $j$, $\forall (i,j) \in \mathcal{I}\times\mathcal{J}$ and $\forall u \in \mathcal{U}$. Since one sub-band is assigned to one association pair, we have
 	\begin{gather} \label{eq:const2}
 		\quad \sum_{i} \sum_{j} z_{j,i}^u = 1,\quad \forall u \in \mathcal{U},
 	\end{gather}
 	and since one IoT is associated with one MEC, we have
 	\begin{gather} \label{eq:const3}
 		\quad \sum_{j} \sum_{u} z_{j,i}^u = 1, \quad \forall i \in \mathcal{I}.
 	\end{gather}
  Owing to the distance and frequency-dependent THz characteristics \cite{ref:Jornet, ref:Han}, the frequency resources need to be allocated adaptively with respect to the communication distance and sub-band frequency to avoid severe molecular absorption loss, thereby attaining high spectral efficiency.

 	\subsection{Task Computation Model}
	As aforesaid, each IoT generates task requests following the Poisson process of rate $\lambda_i$. The task computation system of an MEC server is modeled as a multi-computing unit queue model \cite{ref:Song}, i.e., $M/M/s$ queuing system\footnote{
    In less dynamic environments like IoT sensors in smart farms, the predictable behavior of arrival and service rates allows the transition to an $M/M/s$ model \cite{ref:mmc}, characterized by Poisson arrival processes and exponential service times across multiple servers, thereby facilitating precise performance analysis and optimization of systems with minimal fluctuations.
    }, where $s$ stands for the total number of computing units.
    The \emph{task operation delay} can be modeled as a sum of \emph{queuing delay} and \emph{task computation delay}. The average operation delay of a task arriving at $M/M/s$ queue with given task arrival rate $\lambda$ is 
	\begin{gather} \label{eq:toper} 
		t_\mathsf{oper}(s,\lambda)=\dfrac{C\left(s,\dfrac{\lambda}{\mu}\right)}{s\mu-\lambda}+\dfrac{1}{\mu},
	\end{gather}
	where
	\begin{gather}\label{eq:erlang}
		C(s,\rho)=\dfrac{\left(\dfrac{(s\rho)^s}{s!}\right)\left(\dfrac{1}{1-\rho}\right)}{\sum^{s-1}_{k=0}\dfrac{(s\rho)^k}{k!}+\left(\dfrac{(s\rho)^s}{s!}\right)\left(\dfrac{1}{1-\rho}\right)},
	\end{gather}
	and $\mu$ is the service rate of a computing unit.
	Note that \eqref{eq:erlang} is referred to Erlang's C formula, which depicts the probability that an arriving task is forced to join the queue. To ensure the stability of each edge server queue, the constraint
	\begin{gather}  \label{eq:const5}
	    s > \frac{1}{\mu} \sum_{i} \sum_{u} z_{j,i}^u \cdot \lambda_i , \quad \forall j \in \mathcal{J},
	\end{gather}
	must be satisfied. In other words, \eqref{eq:const5} suggests that the rate of incoming computation requests must not surpass the system's processing capacity to ensure the queue maintains stable operation and efficiently processes all requests.
	
	\subsection{Performance Metric and Problem Formulation}	
	Define \emph{user service delay} by the time elapsed from when a IoT's task request is generated to when the IoT receives the task output that has been processed from an arbitrary server. Then, the user service delay of IoT $i\in \mathcal{I}$ is represented as the sum of \emph{communication delay} $t_{\mathsf{comm},i}$ and \emph{computation delay} $t_{\mathsf{comp},i}$, i.e.,
	\begin{gather*}
		t_{\mathsf{serv},i}=t_{\mathsf{comm},i}+t_{\mathsf{comp},i}.
	\end{gather*}
	Each term is explicitly calculated as follows; first, recall the two types of communication paths, i.e., direct and relay paths. The communication delay of IoT $i$ offloading to MEC $j$ with sub-band $u$ can be expressed as
	\begin{equation*}
		t_{\mathsf{comm},i,j}^u= (1-\alpha_i) t_{\mathsf{direct},i,j}^u		+\sum_{m} \alpha_{i,m}  t_{\mathsf{relay},i,j,m}^{u},
	\end{equation*}
	where the uplink delay through the direct path is obtained as
	\begin{gather*}
        t_{\mathsf{direct},i,j}^u = \frac{D_\mathsf{in}}{R_{i,j}^u},
	\end{gather*}
	and the uplink delay of relay link is obtained as \footnote{A decode-and-forward (DF) relay system operating in a half-duplex mode with adaptive slot length is considered as in \cite{ref:Song, ref:Xu}.}
	\begin{gather*}
		t_{\mathsf{relay},i,j,m}^u = \frac{D_\mathsf{in}}{R_{i,m}^u} + \frac{D_\mathsf{in}}{R_{m,j}^u},
	\end{gather*}  
	where $D_\mathsf{in}$ refers to the task input size. 
	Note that the time spent to download the task output back to IoT over a downlink channel is ignored, since the output data size is relatively  smaller than the input data size, and transmit power of the MEC server is sufficient for the fast transmission \cite{ref:Wu, ref:Fu}, so that the downlink delay is negligibly small compared to that of the uplink. 
    Hence, the overall communication delay of IoT $i \in \mathcal{I}$ can be obtained as
	\begin{gather*}
		t_{\mathsf{comm},i} = \sum_{j} \sum_{u} z_{j,i}^u t_{\mathsf{comm},i,j}^u.
	\end{gather*}
	The uplink delay of each IoT is varying with the user-resource associations for offloading and sub-band, and also the location of UAVs. In particular, because of the distance and frequency-dependent nature of THz bands, these unknown variables jointly affect the uplink delay of entire IoTs, which is complicated and crucial problem.\footnote{In IoT settings like smart farms, frequency bands are allocated specifically for IoT devices, where communication resource constraints are notably less restrictive than those for computing resources. Additionally, the ultra-wideband THz frequency spectrum, providing BW in the order of hundreds of GHz \cite{ref:Shafie}, eliminates communication queue delays by allowing each IoT device's bandwidth to be segmented into multiple bands for multiple requests.
    }
	
	Second, the computation delay of IoT is determined by the average operation delay of the edge server that processes the user's task request. The computation delay of IoT $i \in \mathcal{I}$ can be written as
	\begin{gather*} 
		t_{\mathsf{comp},i}=\sum_{j} \sum_{u} z_{j,i}^u t_\mathsf{oper}\left(s,\sum_{i'}  \sum_{u} z_{j,i'}^u \lambda_{i'}\right),
	\end{gather*}
	where the second argument $\sum_{i'}  \sum_{u} z_{j,i'}^u \lambda_{i'}$ refers to the total task arrival rate at MEC server $j \in \mathcal{J}$, and the task operation delay in (\ref{eq:toper}) is an increasing function with respect to the total task arrival rate. In other words, the computation delay increases as the task requests are overloaded to a server.
	
	In this respect, we aim at minimizing the expected user service delay of overall IoTs by jointly optimizing the UAV relay selection $\bm{\alpha}=[\alpha_{i,m}]_{\mathcal{I} \times \mathcal{M}}$, UAV power control $\mathbf{P}=[P_{i,m}]_{\mathcal{I} \times \mathcal{M}}$, UAV deployment $\mathbf{q}=[\mathbf{q}_{m}]_{\mathcal{M}}$, and user-resource associations $\mathbf{z}=[z_{j,i}^u]_{\mathcal{I} \times \mathcal{J}\times \mathcal{U}}$. 
	Before formulating the problem, we also set a maximum	tolerable user service delay threshold $t_\mathsf{thres}$ for individual IoTs. This is to assure the quality of experience (QoE) of individual	IoTs by preventing any particular user from suffering very	large user service delay, while minimizing the expected user service delay of all the users. The optimization problem for minimizing the overall expected user service delay is formulated as
	\begin{alignat*}{2}
		(\textbf{P1}):\  &\underset{\bm{\alpha}, \mathbf{P}, \mathbf{q}, \mathbf{z}}{\text{min}}	&& 	 \  \frac{1}{I}	\sum_{i} t_{\mathsf{serv},i} \\
		&\text{s.t.}												
        && 	\ 	\eqref{eq:const1},\eqref{eq:const12}, \eqref{eq:const2}, \eqref{eq:const3}, \eqref{eq:const5}, \nonumber \\
		&																	 && 	\ 	\alpha_{i,m}, \in \{0,1\},  \ \forall (i,m) \in \mathcal{I}\times\mathcal{M},\nonumber\\
		&																	 && 	\ 
		z_{j,i}^u \in \{0,1\}, \ \forall (i,j,u) \in \mathcal{I}\times\mathcal{J}\times\mathcal{U}.\nonumber
	\end{alignat*}
    It is important to note that energy consumption is a crucial factor in MEC systems with UAV-enabled communications, as highlighted by several previous studies \cite{ref:Azari2, ref:Zhang3, ref:Mamaghani}. In our considered scenario, we address this factor by incorporating transmit power constraints, namely $P_\text{IoT}$ and $P_\text{UAV}$, into the problem formulation. Our objective is to minimize service delay while operating within these constraints. By doing so, we can effectively optimize the system's performance while considering the limitations of energy consumption. 
    
	In addition, our system primarily focuses on optimizing long-term average performance, especially in scenarios like smart farms or smart factories utilizing IoT and edge computing, where user mobility is limited. In these cases, UAVs deployed remain stationary, resulting in energy consumption primarily divided between transmission and hovering. By constraining transmit power, except for hovering energy, we implicitly control energy usage. Any remaining energy, aside from what's needed for UAV hovering, can be allocated for transmission power. For long-term system viability, it's essential to ensure services while managing finite UAV energy. To achieve this, we categorize UAVs into service and charging groups. This approach segments UAVs for providing services and undergoing recharging. When the service group's batteries are depleted, the charging group can be deployed on-site to implement this strategy.
	
		The problem (\textbf{P1}) is a mixed-integer non-linear programming (MINLP) problem that is generally difficult to solve using existing optimization techniques. Specifically, the search space of binary variables (i.e., $\bm{\alpha}$ and $\mathbf{z}$) grows exponentially with  $I$, $J$, $M$, and $U$, resulting in $2^{I^2JMU}$ possible candidate solutions. 
        Furthermore, all four optimization variables are coupled in the uplink delay, and optimizing the Erlang C formula with the binary variable poses a significant challenge. To overcome the difficulty of handling the combinatorial MINLP problem, we adopt a penalty dual decomposition (PDD)
        -based iterative method \cite{ref:PDD} to effectively optimize (\textbf{P1})\footnote{Due to the inherently static nature of IoT devices, it is feasible to aggregate comprehensive knowledge of system configurations centrally, such as at a control center. This configuration data generally remains stationary, with infrequent changes. Upon establishing the IoT system's configuration, the central unit systematically resolves the optimization of (\textbf{P1}), and enforces the operational policies governing the IoT environment, ensuring the system's overall efficacy to predefined criteria.}.
        
    

    \section{The Proposed Penalty Dual Decomposition-Based Delay Minimization} \label{sec:algorithm}
    
    Due to the presence of binary variables (namely, $\bm{\alpha}$ and $\mathbf{z}$) and the associated challenges found in MINLP problems, directly obtaining the optimal solution for the problem (\textbf{P1}) is nearly impossible. To circumvent this issue, we  employs a double-loop PDD based approach. Initially, we transform the binary constraints into equality constraints by introducing slack variables $\Tilde{\bm{\alpha}}$ and $\Tilde{\mathbf{z}}$:
      \begin{equation}
        \begin{cases}
          \alpha_{i,m} (\Tilde{\alpha}_{i,m}-1)=0, & \forall (i, m) \in \mathcal{I} \times \mathcal{M}, \\
          \alpha_{i,m}  -\Tilde{\alpha}_{i,m}=0, &  \forall (i, m) \in \mathcal{I} \times \mathcal{M},
        \end{cases}  \label{cons:pdda}
      \end{equation}
    and
      \begin{equation}
        \begin{cases}
          z_{j,i}^u (\Tilde{z}_{j,i}^u-1)=0, & \forall (i,j,u) \in \mathcal{I} \times \mathcal{J} \times \mathcal{U}, \\
          z_{j,i}^u  -\Tilde{z}_{j,i}^u=0, &  \forall (i,j,u) \in \mathcal{I} \times\mathcal{J} \times  \mathcal{U}. 
        \end{cases} \label{cons:pddz}
      \end{equation}
    Subsequently, the original problem (\textbf{P1}) can be reformulated into an augmented Lagrangian (AL) problem by dualizing and penalizing the equality constraints, specifically \eqref{eq:const2}, \eqref{eq:const3}, \eqref{cons:pdda}, and \eqref{cons:pddz}, utilizing the penalized parameters $\rho_\alpha$ and $\rho_z$. This is achieved as outlined in the following.
	\begin{alignat*}{2}
		(\textbf{P2}):\  &\underset{\bm{\alpha}, \Tilde{\bm{\alpha}}, \mathbf{P}, \mathbf{q}, \mathbf{z}, \Tilde{\mathbf{z}} }{\text{min}}	&& 	 \  	\sum_{i} t_{\mathsf{serv},i} 
  +  \frac{1}{2\rho_\alpha} \Lambda_\alpha +  \frac{1}{2\rho_z} \Lambda_z
  \\
		&\text{s.t.}						
        && 	\ 	\eqref{eq:const1},\eqref{eq:const12},  \eqref{eq:const5}.\nonumber 
	\end{alignat*}
    Here, $\Lambda_\alpha = \sum_i \sum_m \Lambda_{{i,m}}^\alpha$ and $\Lambda_z=\sum_j \sum_i \sum_u \Lambda_{j,i,u}^z + \sum_u \Lambda_{u}^z + \sum_i \Lambda_{i}^z$, where 
    \begin{alignat*} {2}
    &\Lambda_{{i,m}}^\alpha =  \big| \alpha_{i,m} (\Tilde{\alpha}_{i,m}-1)  + \rho_\alpha &&\eta^{\alpha, 1}_{i,m}\big|^2  
    \\&   &&
     +\big| \alpha_{i,m}  -\Tilde{\alpha}_{i,m}  + \rho_\alpha\eta^{\alpha,2}_{i,m}\big|^2, 
    \end{alignat*}  
    \begin{equation*}
    \Lambda_{j,i,u}^z =  \big| z_{j,i}^u (\Tilde{z}_{j,i}^u-1)  +\rho_z  \eta^{z,1}_{j,i,u}\big|^2 + \big| z_{j,i}^u  -\Tilde{z}_{j,i}^u + \rho_z\eta^{z,2}_{j,i,u} \big|^2,
    \end{equation*}  
    \begin{equation*}
    \Lambda_{u}^z =  \bigg| \sum_j \sum_i z_{j,i}^u -1  +\rho_z\eta^{z}_{u}\bigg|^2, 
    \Lambda_{i}^z =  \bigg| \sum_j \sum_u z_{j,i}^u -1 +\rho_z\eta^{z}_{i}\bigg|^2, 
    \end{equation*}  
    and $\{\eta^{\alpha,1}_{i,m},\eta^{\alpha,2}_{i,m}, \eta^{z,1}_{j,i,u}, \eta^{z,2}_{j,i,u}, \eta^{z}_{u}, \eta^{z}_{i}\}$ are  dual vairables. It is important to note that the problem (\textbf{P2}) becomes equivalent to the original problem (\textbf{P1}) as the penalized parameters $\rho_\alpha$ and $\rho_z$ approach zero.

    The proposed method incorporates a dual-loop structure to optimize (\textbf{P2}). Within the inner loop, with penalty parameters and  dual vairables held constant and (\textbf{P2}) is divided into four distinct sub-problems: UAV relay selection, UAV power control, UAV deployment, and user-resource association optimization, each addressed separately. Conversely, in the outer loop, both the penalized parameters ${\rho_\alpha, \rho_z}$ and the  dual vairables $\{\eta^{\alpha,1}_{i,m},\eta^{\alpha,2}_{i,m}, \eta^{z,1}_{j,i,u}, \eta^{z,2}_{j,i,u}, \eta^{z}_{u}, \eta^{z}_{i}\}$ undergo  updates. The specific procedures and methodologies  are elaborated in subsequent subsections.  Despite the PDD iterative optimization method treating sub-problems independently, it substantially lowers computational complexity while achieving near-optimal solutions. This makes it highly practical for complex, high-dimensional optimization tasks where joint optimization is computationally infeasible.

	\subsection{Inner Loop: UAV Relay Selection}
	When the UAV location $\mathbf{q}$, UAV power control $\mathbf{P}$, and user-resource associations $\mathbf{z}, \Tilde{\mathbf{z}}$ are given, the optimization problem (\textbf{P2}) reduces to
	\begin{alignat*}{2}
		(\textbf{SP1}): \ &\underset{\bm{\alpha}, \Tilde{\bm{\alpha}}}{\text{min}}	&& 	 \sum_{i} t_{\mathsf{comm},i} +  \frac{1}{2\rho_\alpha} \Lambda_\alpha \\
		&\text{s.t.}
            && 	\sum_{m} \alpha_{i, m} \leq 1, \ \forall i \in \mathcal{I},
	\end{alignat*}
	where $t_{\mathsf{comm},i} = (1-\alpha_i) t_{\mathsf{direct},i} + \sum_m \alpha_{i,m} t_{\mathsf{relay},i,m}$ and $t_{\mathsf{relay},i,m} = \sum_j \sum_u z_{j,i}^u t_{\mathsf{relay},i,j,m}^u$. 
	Since the UAV relay selection is decided by users independently, (\textbf{SP1}) can be further divided into per-user optimization for user $i$ as
	\begin{alignat}{2} \label{opt:perusersp1}
		 \ &\underset{{\alpha}_i, \Tilde{{\alpha}}_i}{\text{min}}	 	  && ~ t_{\mathsf{comm},i} +  \frac{1}{2\rho_\alpha}  \Lambda_{i}^\alpha \\
		&\text{s.t.} 
            && 	\sum_{m} \alpha_{i, m} \leq 1,  \nonumber
	\end{alignat}
    where $\Lambda_{i}^\alpha=\sum_m \big\{ | \alpha_{i,m} (\Tilde{\alpha}_{i,m}-1)  + \rho_\alpha \eta^{\alpha, 1}_{i,m}|^2   + | \alpha_{i,m}  -\Tilde{\alpha}_{i,m}  + \rho_\alpha\eta^{\alpha,2}_{i,m}|^2 \big\}$. 
    Since the \eqref{opt:perusersp1} is convex in terms of $\Tilde{{\alpha}}_{i, m}$, by solving $\frac{\partial\Lambda_{i}^\alpha}{\partial\Tilde{{\alpha}}_{i, m}}=0$, the closed-form solution of $\Tilde{{\alpha}}_{i, m}$ to \eqref{opt:perusersp1} is given by
        \begin{equation*}
            \Tilde{{\alpha}}_{i, m}^* = \frac{\alpha_{i, m}^2+(1-\rho_\alpha\eta^{\alpha,1}_{i,m})\alpha_{i, m} + \rho_\alpha\eta^{\alpha,2}_{i,m}}{\alpha_{i, m}^2+1} 
        \end{equation*}
  Accordingly, using the result above,\eqref{opt:perusersp1} can be rewritten as
	\begin{alignat}{2} \label{opt:peruser2sp1}
		 \ &\underset{{\alpha}_i}{\text{min}}	 	  && ~ t_{\mathsf{comm},i} +  \frac{1}{2\rho_\alpha}  {\Lambda_{i}^\alpha}^\prime \\
		&\text{s.t.} 
            && 	\sum_{m} \alpha_{i, m} \leq 1, \nonumber
	\end{alignat}
  where ${\Lambda_{i}^\alpha}^\prime=\sum_m \big\{ | \alpha_{i,m} (\Tilde{\alpha}_{i,m}^*-1)  + \rho_\alpha \eta^{\alpha, 1}_{i,m}|^2   + | \alpha_{i,m}  -\Tilde{\alpha}_{i,m}^*  + \rho_\alpha\eta^{\alpha,2}_{i,m}|^2 \big\}$. After several straightforward derivation steps, it becomes evident that the problem is convex. Consequently, it can be efficiently solved using convex optimization tools such as CVX \cite{ref:CVX} and YALMIP \cite{ref:YALMIP}. 
  
  Nonetheless, we introduce the following theorem to present the closed-form solution of (\textbf{SP1}) that can be obtained when the algorithm ultimately reaches convergence, i.e., $\rho_\alpha \rightarrow 0$.

  \begin{theorem} \label{thm:sp1}
    The optimal solution for (\textbf{SP1}) is given by
    \begin{equation*}
     \alpha_{i,m}^* = 
      \begin{cases}
          1, \text{if } m =\argmin_{m^\prime} t_{\mathsf{relay},i,m^\prime} \text{ and }  t_{\mathsf{direct},i} > t_{\mathsf{relay},i,m} \\
          0, \text{otherwise},
      \end{cases}  
    \end{equation*}
    for all $i \in \mathcal{I}$, as $\rho_\alpha \rightarrow 0$.
    \end{theorem}
    \begin{IEEEproof}
        After plugging $\Tilde{{\alpha}}_{i, m}^*$ into ${\Lambda_{i}^\alpha}^\prime$ and some derivations, we can rewrite ${\Lambda_{i}^\alpha}^\prime$ as
        \begin{equation*}
        {\Lambda_{i}^\alpha}^\prime = \sum_m \bigg\{ \frac{(\alpha_{i,m}^2 +(\rho_\alpha \eta^{\alpha,2}_{i,m} -1)\alpha_{i,m}+\rho_\alpha \eta^{\alpha,1}_{i,m} )^2}{1+\alpha_{i,m}^2} \bigg\}
        \end{equation*}
        As $\rho_\alpha \rightarrow 0$, 
        we have the minimization of 
	\begin{alignat*}{2} 
		 \ &\underset{{\alpha}_i}{\text{min}}	 	  && ~ \sum_m \bigg\{ \alpha_{i,m} (t_{\mathsf{relay},i,m}-t_{\mathsf{direct},i}) + \lim_{\rho_\alpha \rightarrow0}\frac{1}{2\rho_\alpha}{\Lambda_{i}^\alpha}^\prime \bigg\} \\
		&\text{s.t.} 
            && 	\sum_{m} \alpha_{i, m} \leq 1.
	\end{alignat*}
        The penalty term becomes 
        \begin{alignat*}{2}
            \lim_{\rho_\alpha \rightarrow0}\frac{1}{2\rho_\alpha}{\Lambda_{i}^\alpha}^\prime=&\frac{\alpha_{i,m}(\alpha_{i,m}-1)(\eta^{\alpha,1}_{i,m}+\eta^{\alpha,2}_{i,m}\alpha_{i,m})}{2(1+\alpha_{i,m}^2)} \\&+ \lim_{\rho_\alpha \rightarrow0}\frac{1}{2\rho_\alpha}\frac{\alpha_{i,m}^2(\alpha_{i,m}-1)^2}{1+\alpha_{i,m}^2},
        \end{alignat*}
        implying that the penalty term diverges if $\alpha_{i,m}$ is neither 0 nor 1. Therefore, in terms of minimizing 
        $\sum_m \alpha_{i,m} (t_{\mathsf{relay},i,m}-t_{\mathsf{direct},i})$        while also ensuring that at most one UAV is selected for IoT $i$, each IoT $i \in \mathcal{I}$ selects a UAV only when the minimum communication delay of the relay path among UAVs is smaller than that of the direct path, i.e., $t_{\mathsf{direct},i}>\min_m t_{\mathsf{relay},i,m}$. Thus, for all $i \in \mathcal{I}$, we have  
        \begin{equation*}
             \alpha_{i,m}^* = 
              \begin{cases}
                 1, \text{if } m =\argmin_{m^\prime} t_{\mathsf{relay},i,m^\prime} \text{ and }  t_{\mathsf{direct},i} > t_{\mathsf{relay},i,m} \\
          0, \text{otherwise},
             \end{cases}  
        \end{equation*} 
    \end{IEEEproof}


	\subsection{Inner Loop: UAV Power Control}
	Given the UAV relay selection $\bm{\alpha}, \Tilde{\bm{\alpha}}$, UAV positioning $\mathbf{q}$, and user-resource associations $\mathbf{z}, \Tilde{\mathbf{z}}$, the original optimization problem (\textbf{P2}) is reformulated as 
	\begin{alignat}{2}
		(\textbf{SP2}): \ &\underset{\mathbf{P}}{\text{min}}	&&  	\sum_{i} t_{\mathsf{comm},i} \label{eq:sp2obj} \\
		&\text{s.t.}	
                && 	\sum_{i} P_{i, m} \leq P_\mathsf{UAV}, \ \forall m\in \mathcal{M}, \nonumber
	\end{alignat}
	where the UAV power control $\mathbf{P}$ is also independent of the computation delay but only has a dependency on the communication delay. 

After several straightforward derivations, it can be easily verified that the Hessian matrix of the objective function \eqref{eq:sp2obj} is positive definite, confirming that the objective function is convex with respect to the power control variable $\mathbf{P}$.

	Hence, the subproblem (\textbf{SP2}) is a convex problem, which can be efficiently solved via well-known convex optimization tools such as CVX \cite{ref:CVX} and YALMIP \cite{ref:YALMIP}.
 
    In addition, as the algorithm converges and $z$ approaches binary values, the optimal solution structure of (\textbf{SP2}) can be elucidated as follows.  Using the Karush-Kuhn-Tucker (KKT) condition of (\textbf{SP2}), we can obtain the optimal power control. 

    \begin{theorem} \label{thm:sp2}
    The optimal solution for (\textbf{SP2}) is given by
    \begin{equation*}
     P^*_{i,m} = \frac{1}{\gamma_{i,m}}\bigg( e^{2W\big(\frac{1}{2} \sqrt{\frac{l_{i,m}}{\mu_m}}\big)} - 1\bigg) ,  
    \end{equation*}
    where $\gamma_{i,m} = \frac{|h_{m,j}^u|^2}{BN_0}$, $l_{i,m} = \frac{D_\text{in} \ln{2} }{B}\alpha_{i,m} \gamma_{i,m}\sum_j \sum_u z_{j,i}^u  $, $\mu_m$ is a Lagrangian multiplier for UAV $m$, and $W(\cdot)$ is the Lambert W function
    , as $\rho_z \rightarrow 0$.
    \end{theorem}
    \begin{IEEEproof} 
    The Lagrangian dual function of (\textbf{SP2}) is 
    \begin{equation*}
        \mathcal{L}(\mathbf{q}, \mu) =\sum_i t_{\text{comm}, i} +\sum_m \mu_m (\sum_i P_{i,m} - P_\text{UAV}),
    \end{equation*}
    where $\mu_m$ is a Lagrangian multiplier for UAV $m$. 
    
    By KKT condition, we have 
    \begin{equation*}
    \begin{cases}
        \frac{\partial \mathcal{L}}{\partial P_{i,m}}=\mu_m -  \frac{l_{i,m}}{(\ln{(1+\gamma_{i,m}P_{i,m})})^2 (1+\gamma_{i,m}P_{i,m})}=0, \forall i, m,    \\
        \sum_i P_{i,m} \leq P_\text{UAV}, \forall m, \\
        \mu_m \geq 0, \forall m, \\
        \mu_m (\sum_i P_{i,m} - P_\text{UAV})=0, \forall m,
    \end{cases}
    \end{equation*}
    where $\gamma_{i,m} = \frac{|h_{m,j}^u|^2}{BN_0}$, $l_{i,m} = \frac{D_\text{in} \ln{2} }{B}\alpha_{i,m} \gamma_{i,m}\sum_j \sum_u z_{j,i}^u   $.
    To satisfy the complementary slackness (i.e., $\mu_m (\sum_i P_{i,m} - P_\text{UAV})=0$), we have two cases as below. 
    
    \emph{Case (1) $\mu_m = 0$} : From the stationarity condition, we have
    \begin{equation*}
        \frac{l_{i,m}}{(\ln{(1+\gamma_{i,m}P_{i,m})})^2 (1+\gamma_{i,m}P_{i,m})}=0,
    \end{equation*}
    where the solution for this equation becomes $P_{i,m}=\infty$, which is infeasible. 
    
    \emph{Case (2) $\mu_m \neq 0$ and $\sum_i P_{i,m} - P_\text{UAV} = 0$} : From the stationarity condition, we have
    \begin{alignat*} {2}
        &\frac{l_{i,m}}{(\ln{(1+\gamma_{i,m}P_{i,m})})^2 (1+\gamma_{i,m}P_{i,m})}=\mu_m \\
        \leftrightarrow ~ &\frac{l_{i,m}}{\mu_m}=\bigg(\ln{(1+\gamma_{i,m}P_{i,m})} e^{\frac{\ln{(1+\gamma_{i,m}P_{i,m})}}{2}}\bigg)^2 \\
        \leftrightarrow ~ & \ln{(1+\gamma_{i,m}P_{i,m})}= 2 \cdot W\bigg(\frac{1}{2} \sqrt{\frac{l_{i,m}}{\mu_m}}\bigg), 
    \end{alignat*} 
    where $W(\cdot)$ is the Lambert W function. From this aspect, the feasible power control can be obtained by
    \begin{equation} \label{eq:optpower}
     P_{i,m} = \frac{1}{\gamma_{i,m}}\bigg( e^{2W\big(\frac{1}{2} \sqrt{\frac{l_{i,m}}{\mu_m}}\big)} - 1\bigg).
    \end{equation}
    Then, we should show that the feasible solution \eqref{eq:optpower} can satisfy $\sum_i P_{i,m} = P_\text{UAV}$. We know that $\frac{l_{i,m}}{\mu_m}>0$, thus $W\big(\frac{1}{2} \sqrt{\frac{l_{i,m}}{\mu_m}}\big)>0$, which is decreasing with respect to $\mu_m$. Therefore, we can always determine $\{\mu_m, \forall m\in \mathcal{M}\}$ such that $\sum_i \frac{1}{\gamma_{i,m}}\big( e^{2W\big(\frac{1}{2} \sqrt{\frac{l_{i,m}}{\mu_m}}\big)} - 1\big) = P_\text{UAV}$ and $\mu_m>0$, which can be obtained via the bisection search, and \eqref{eq:optpower} is the feasible optimal solution.
    \end{IEEEproof}

	\subsection{Inner Loop: UAV Positioning}
	For fixed UAV relay selection $\bm{\alpha}, \Tilde{\bm{\alpha}}$, UAV power control $\mathbf{P}$, and user-resource associations $\mathbf{z}, \Tilde{\mathbf{z}}$, the optimization problem (\textbf{P1}) can be recast as \begin{alignat}{2}
		(\textbf{SP3}): \ &\underset{\mathbf{q}}{\text{min}} && \sum_{i} t_{\mathsf{comm},i} \label{eq:sp3obj} 
	\end{alignat}
	It is noted that the UAV placements affect the communication distances of IoT-to-UAV and UAV-to-MEC server, not the distance of the IoT-to-MEC server. In other words, the UAV placements solely affect the relay path's uplink delay, not the direct path. Thus, we focus on minimizing the relay uplink delay.
	To solve the subproblem (\textbf{SP3}), we need to first prove the convexity of \eqref{eq:sp3obj}. 
 The following lemma states the convexity of \eqref{eq:sp3obj}.
	\begin{lemma} \label{thm1}
		The objective function in \eqref{eq:sp3obj} is convex with respect to the UAV placement $\mathbf{q}$ under the condition
		\begin{align}\label{eq:convexcondition}
		d(\mathbf{q}_m) \! > \! \frac{1}{K}\!\left(\sqrt{\frac{2g(\mathbf{q}_m)\ln{g(\mathbf{q}_m)}}{2g(\mathbf{q}_m)\!-\!2\!-\!\ln{g(\mathbf{q}_m)}}}\!-\!2\right),\ \forall m \!\in\! \mathcal{M}.
		\end{align}
		where $K$ represents the molecular absorption coefficient and $g(\mathbf{q}_m) = 1 + \frac{P |h(\mathbf{q}_m)|^2}{BN_0}$. 
	\end{lemma}
	\begin{IEEEproof}
		To begin with, note that the objective function in \eqref{eq:sp3obj} is equivalent to $\sum_{i} \sum_{m } \alpha_{i,m} D_\mathsf{in} \frac{ \ln2}{B} \cdot (t_{\mathbf{(SP3)}, 1} (\mathbf{q}_m) + t_{\mathbf{(SP3)}, 2}(\mathbf{q}_m))$ for fixed variables $\alpha$, $\mathbf{P}$, and $\mathbf{z}$, where $\frac{ \ln2}{B} t_{\mathbf{(SP3)}, 1}$ and $ \frac{ \ln2}{B} t_{\mathbf{(SP3)}, 2}$ refer to a reciprocal of the data rate of IoT-to-UAV and UAV-to-MEC, respectively. Since the sum of convex functions is also convex, let us focus on proving the convexity of a function $t_\mathbf{(SP3)}(\mathbf{q}_m) = \frac{1}{\ln(g(\mathbf{q}_m))}$, where $g(\mathbf{q}_m) = 1 + \frac{P |h(\mathbf{q}_m)|^2}{BN_0}$.
		
		The channel gain can be rewritten as 
		\begin{gather*}
			|h(\mathbf{q}_m)|^2=\left(\frac{s_\text{light}}{4\pi f d(\mathbf{q}_m)}\right)^2 e^{-K(f)d(\mathbf{q}_m)}
		\end{gather*} 
		where $d(\mathbf{q}_m)= \sqrt{(x_m - x)^2 + (y_m - y)^2 + H^2}$. Here, we set the coordinates of the IoT or MEC server that communicates with the UAV $m$ as $[x, y, 0]$ for the ease of expression. 
		
		To prove the convexity, we need to prove whether the principal minors of the Hessian matrix are positive or not. Hereinafter, $t_\mathbf{(SP3)}(\mathbf{q}_m)$, $g(\mathbf{q}_m)$, $d(\mathbf{q}_m)$, and $K(f)$ are shortened to $t_\mathbf{(SP3)}$, $g$, $d$, and $K$ for the sake of brevity. The first-order principal minor of $x_m$ (or $y_m$) can be deployed as 
		\begin{gather*}
			\frac{\partial^2 t_\mathbf{(SP3)}}{\partial x_m^2} =  \frac{g-1}{d^2 g^2 \ln^3{g}} Q,
		\end{gather*}
		where $Q =  g \ln{g} (2+Kd)d^2   + (x_m - x)^2 \{(2g-2-\ln{g})(2+Kd)^2-g \ln{g} (2+Kd)-2g \ln{g}\}$.	Since $g>1$, we now concentrate on proving the positivity of $Q$. 
		First, it is obvious that if $(2g-2-\ln{g})(2+Kd)^2-g \ln{g} (2+Kd)-2g \ln{g} \geq 0$, then $Q > 0$. If $(2g-2-\ln{g})(2+Kd)^2-g \ln{g} (2+Kd)-2g \ln{g} < 0$
		, since $d^2 \geq (x_m - x)^2$, the lower bound of $Q$ can be obtained as
		\begin{align}
			\nonumber Q &\geq d^2  [g \ln{g} (2+Kd)-g \ln{g} (2+Kd)\\ & \quad\quad\quad\quad  + (2g-2-\ln{g})(2+Kd)^2-2g \ln{g}] \nonumber\\
			&= d^2 [(2g-2-\ln{g})(2+Kd)^2-2g \ln{g}]\label{eq:Qlower}
		\end{align}
		Thus, if the right-hand-side (RHS) of \eqref{eq:Qlower} is positive, then so is $Q$. Since $g > 1$ from the definition, $2g-2-\ln{g} > 0$, and thus the condition for the RHS of \eqref{eq:Qlower} being positive is \eqref{eq:convexcondition}.
		The procedures on $y_m$ is the same.
		To shed light on this condition \eqref{eq:convexcondition} of convexity, we introduce some notable intuitions, which will be discussed in Remark \ref{rem:1}.
		 
		Next, we prove the positivity of the second-order principal minor, which can be represented as
		\begin{gather*}
			\begin{vmatrix}
				\frac{\partial^2 t_\mathbf{(SP3)}}{\partial x_m^2}  &\frac{\partial^2 t_\mathbf{(SP3)}}{\partial x_m \partial y_m}  \\ 
				\frac{\partial^2 t_\mathbf{(SP3)}}{\partial y_m \partial x_m} & \frac{\partial^2 t_\mathbf{(SP3)}}{\partial y_m^2} 
			\end{vmatrix} 
		= \frac{(g-1)^2}{d^6 g^3 \ln^3{g}} W,
		\end{gather*}
		where $W=H^2(2+Kd)g\ln{g}+(d^2-H^2)\{(2g-2-\ln{g})(2+Kd)^2-2g\ln{g}\}$. From the fact that $Kd \geq 0$, $W$ is lower bounded as
		\begin{equation}\label{eq:second_order_principal}
			\begin{alignedat}{2}
			W \! \geq \! 2H^2g\ln{g}\!+\!(d^2\!-\!H^2)\{8g\!-\!8\!-\!4\ln{g}\!-\!2g\ln{g}\}. 
		\end{alignedat}
		\end{equation}
		Note that if $(2g-2-\ln{g})(2+0)^2-2g\ln{g}$ is positive, then the RHS of  \eqref{eq:second_order_principal} is positive. Thus, if the condition \eqref{eq:convexcondition} is satisfied, the second-order principal minor is positive.
		
		On this account, $t_\mathbf{(SP3)}(\mathbf{q}_m)$ is convex with respect to $\mathbf{q}_m$ when satisfying \eqref{eq:convexcondition}, and also the variables $\mathbf{q}_m, \forall m \in \mathcal{M}$ independently influence to the objective function \eqref{eq:sp3obj}, thus we can conclude that the objective function \eqref{eq:sp3obj} is convex with respect to the UAV placement variable $\mathbf{q}$ when satisfying the condition \eqref{eq:convexcondition}.
	\end{IEEEproof}
    
	Accordingly, (\textbf{SP3}) is also a convex problem, which can be solved by standard convex optimization tools.
	
\begin{figure}[t]
		{\label{fig:uplink}\includegraphics[width=0.7\columnwidth]{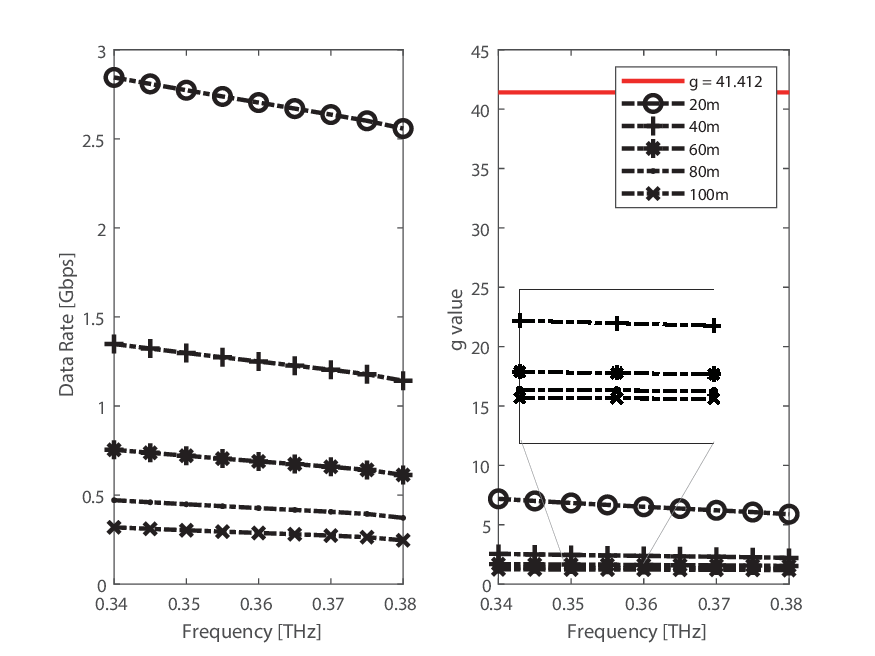}} 
		\centering
		\caption{Comparison of uplink data rate and $g$ value with respect to different communication distances with $B=1$GHz and $P=2$W.}
		\vspace{-10pt}
		\label{fig:gval}
	\end{figure}

	\begin{remark} \label{rem:1}
        Since $Kd \geq 0$, the condition \eqref{eq:convexcondition} can be relaxed to $\sqrt{\frac{2g \ln{g}}{2g-2-\ln{g}}} < 2$, which is satisfied when the signal-to-noise ratio (SNR) $g < 41.412$. Note that $g$ increases with transmit power and channel path gain but decreases with bandwidth. In the THz band, this condition is easily met with wide-band usage, as shown in Fig. \ref{fig:gval}. The figure illustrates uplink data rates and $g$ values for different communication distances in the $[0.34, 0.38]$ THz band with $B = 1$GHz and $P = 2$W as  \cite{ref:Mamaghani, ref:Xu, ref:Pan, ref:Hassan, ref:Shafie}. These results show that $g$ is generally less than $8$, much smaller than $41.412$, while providing sufficient uplink data rates. Therefore, we can conclude that the convexity of (\textbf{SP3}) is generally satisfied in the THz band.
    \end{remark}
	
	\subsection{Inner Loop: User-resource associations}
	Given the UAV selection $\bm{\alpha}, \Tilde{\bm{\alpha}}$, UAV power control $\mathbf{P}$, and the UAV positioning $\mathbf{q}$, the problem (\textbf{P1}) reduces to the user-resource associations optimization problem as 
	\begin{alignat}{2}
		(\textbf{SP4}): \quad &\underset {\mathbf{z}, \Tilde{\mathbf{z}}}{\text{min}}\sum_{i} \left( t_{\mathsf{comm},i} + t_{\mathsf{comp},i} \right) + \frac{1}{2\rho_z} \Lambda_z \nonumber\\ 
		&\text{s.t.} 
		 	~\eqref{eq:const2}, \eqref{eq:const3}  \text{ and } \eqref{eq:const5},\nonumber
	\end{alignat}
	Note that the major difficulty of resolving the subproblem (\textbf{SP4}) is in  handling the Erlang C formula in $t_{\mathsf{comp},i}$, that is extremely hard. To  alleviate the colossal complexity of Erlang C formula, we apply the tight and tractable upper bound\footnote{The tightness of this upper bound will be discussed in Remark \ref{rem:conv}.} proposed in \cite{ref:Harel}, which is
	\begin{gather} \label{eq:uberlang}
		\quad \quad C\left(s, \frac{\lambda}{\mu}\right) < \left( \frac{\lambda}{s\mu} \right)^{\sqrt{s}}, \quad s \geq 2.
	\end{gather}
	Notice that it is possible to always satisfy  $s\geq 2$ by dividing computing units into more than two virtual machines with equal computing ability \cite{ref:Cheng}.
	Considering this, the task operation delay \eqref{eq:toper} can be rewritten as
	\begin{gather*} 
		t_\mathsf{oper}(s,\lambda) < \frac{\big(\lambda / {s\mu}  \big)^{\sqrt{s}}}{s\mu-\lambda}+\dfrac{1}{\mu} \triangleq \overline{t}_\mathsf{oper}(s,\lambda).
	\end{gather*}
	Consequently, we have
	\begin{gather*} 
		t_{\mathsf{comp},i} < \sum_{j} \sum_{u} z_{j,i}^u  \overline{t}_\mathsf{oper}\left(s,\sum_{i'} \sum_{u}  z_{j,i'}^u \lambda_{i'}\right) \triangleq \overline{t}_\mathsf{comp,i}.
	\end{gather*}

	On balance, the subproblem (\textbf{SP4}) can be reformulated as
	\begin{alignat}{2}
		(\textbf{SP4.1}): \ &\underset{\mathbf{z}, \Tilde{\mathbf{z}}}{\text{min}}\sum_{i} \left( t_{\mathsf{comm},i} + \overline{t}_{\mathsf{comp},i} \right) + \frac{1}{2\rho_z} \Lambda_z \label{eq:sp42obj2}\\
			&\text{s.t.} \ ~
            \eqref{eq:const5}. \nonumber
	\end{alignat}
Subsequently, (\textbf{SP4.1}) is partitioned into two separate optimization blocks, designated for $\mathbf{z}$ and $\Tilde{\mathbf{z}}$, respectively.

\emph{(1) Updating $\mathbf{z}$} :
To initiate the optimization of $\mathbf{z}$, we observe that ${t}_{\mathsf{comm},i}$ exhibits linearity, while the penalty term embodies a convex function with respect to $z_{j,i}^u$. For the purpose of optimizing $\mathbf{z}$ through convex optimization methods, it becomes imperative to ensure that $\overline{t}_{\mathsf{comp},i}$ is rendered convex, signifying that the upper-bounded objective $\overline{t}_{\mathsf{serv},i}$ also adopts a convex form. In light of this requirement, we introduce the following lemma to unveil the convexity of $\overline{t}_{\mathsf{comp},i}$.
	
	\begin{lemma}
		$\overline{t}_{\mathsf{comp},i}$ is a convex function with respect to $z_{j,i}^u$.
	\end{lemma}
	\begin{IEEEproof}
	    To offer the proof of convexity, we first provide the proof that $\overline{t}_\mathsf{oper}(s,\lambda)$ is convex with respect to $\lambda$, where $\lambda$ is an affine function of $z_{j,i}^u$. 
	    
	    The second-order derivative of $\overline{t}_\mathsf{oper}(s,\lambda)$ is
	    \begin{gather*}
	    	\frac{\partial^2 \overline{t}_\mathsf{oper}(s,\lambda)}{\partial \lambda^2}=\frac{\lambda^{\sqrt{s}-2}}{(s\mu - \lambda)^3}  E,
	    \end{gather*}
	    where $E=(2-\sqrt{s})(1-\sqrt{s})\left(\lambda^2+\frac{2\mu s^{3/2}}{1-\sqrt{s}}\lambda + \frac{\mu^2 s^{5/2}}{\sqrt{s}-2}\right)$, for $\lambda<s\mu$ and $s\geq2$. Then, we need to prove whether $E$ is positive or not. We have two cases; (1) $\sqrt{s} < 2$ and (2) $\sqrt{s} > 2$. Note that $E = 0$ when $\sqrt{s} = 2$.
	    
	    \emph{Case (1) $\sqrt{s} < 2$} : Since $(2-\sqrt{s})(1-\sqrt{s})$ is negative, $\frac{E}{(2-\sqrt{s})(1-\sqrt{s})}$ should also be negative. Note that $\lambda^2+\frac{2\mu s^{3/2}}{1-\sqrt{s}}\lambda + \frac{\mu^2 s^{5/2}}{\sqrt{s}-2}$ can be seen as a quadratic equation of $\lambda$, and its discriminant is 
	    \begin{gather}
	    	\frac{\mu^2 s^3}{\sqrt{s}(1-\sqrt{s})^2 (2-\sqrt{s})}, \label{eq:disc}
	    \end{gather}
    which is positive. Thus, there are two $\lambda$-axis intercepts, which are $\lambda_+ = \frac{\mu s^{3/2}}{\sqrt{s}-1}(1 + (\sqrt{s}(2-\sqrt{s}))^{-\frac{1}{2}})$ and $\lambda_- = \frac{\mu s^{3/2}}{\sqrt{s}-1}(1 - (\sqrt{s}(2-\sqrt{s}))^{-\frac{1}{2}})$. Since $0<\lambda<s\mu$, the two intercepts should respectively satisfy $\lambda_- < 0$ and $\lambda_+ > s\mu$ to make $\frac{E}{(2-\sqrt{s})(1-\sqrt{s})}$ negative. Since $\frac{\mu s^{3/2}}{\sqrt{s}-1} > 0$ and $1- (\sqrt{s}(2-\sqrt{s}))^{-\frac{1}{2}} < 0$, $\lambda_- < 0$ holds. On the other hand, since $\frac{\sqrt{s}}{\sqrt{s}-1}>1$ and $1+ (\sqrt{s}(2-\sqrt{s}))^{-\frac{1}{2}} > 1$, $\lambda_+$ holds. Therefore, $\frac{E}{(2-\sqrt{s})(1-\sqrt{s})}$ is negative and $E$ is positive.

	\emph{Case (2) $\sqrt{s} > 2$} : Similarly, since $(2-\sqrt{s})(1-\sqrt{s})$ is positive, $\frac{E}{(2-\sqrt{s})(1-\sqrt{s})}$ should be positive. In this case, its discriminant \eqref{eq:disc} is negative, so that the quadratic function $\frac{E}{(2-\sqrt{s})(1-\sqrt{s})}$ is always above the $\lambda$-axis. Thus, $E$ is positive when $\sqrt{s}>2$.
		
	Put together, the second derivative of $\overline{t}_\mathsf{oper}(s,\lambda)$ is  always positive, implicating that the function $\overline{t}_\mathsf{oper}(s,\lambda)$ is convex with respect to $\lambda$. Besides, the composition of the convex function with affine mapping (i.e., $\overline{t}_\mathsf{oper}(s,\sum_{i^{'}} \sum_{u} z_{j,i^{'}}^u \lambda_{i^{'}})$) and the product of linear and increasing convex function\footnote{The operation delay $\overline{t}_\mathsf{oper}(s,\sum_{i^{'}} \sum_{u} z_{j,i^{'}}^u \lambda_{i^{'}})$ is increasing as the task request arrivals increase.} (i.e., $\sum_j \sum_{u} z_{j,i}^u \overline{t}_\mathsf{oper}(s,\sum_{i^{'}} \sum_{u} z_{j,i^{'}}^u \lambda_{i^{'}})$) are also known as a convex \cite{ref:boyd}, we conclude that $\overline{t}_{\mathsf{comp},i}$  is a convex function.
	\end{IEEEproof}
	As a result, $\mathbf{z}$ can be effectively optimized by using the standard convex optimization solvers such as CVX and YALMIP. 

 \emph{(2) Updating $\Tilde{\mathbf{z}}$} : 
    Then, the slack variable $\Tilde{\mathbf{z}}$ can be updated by the closed-form solution for $\Tilde{z}_{j,i}^u$ in \eqref{eq:sp42obj2} for a given $z_{j,i}^u$:
        \begin{equation*}
            (\Tilde{z}_{j,i}^{u})^* = \frac{(z_{j,i}^u)^2+(1-\rho_z\eta^{z,1}_{j,i,u})z_{j,i}^u + \rho_z \eta^{z,2}_{j,i,u}}{(z_{j,i}^u)^2+1} 
        \end{equation*}
    This is because that the \eqref{eq:sp42obj2} is convex with respect to $\Tilde{{z}}_{j,i}^u$, it can simply be derived by solving $\frac{\partial\Lambda_{z}}{\partial\Tilde{z}_{j,i}^u}=0$.

	\subsection{Outer Loop: Dual variables and penalized parameters} 
        In the outer loop, the  dual vairables $\bm{\eta}=\{\eta^{\alpha,1}_{i,m},\eta^{\alpha,2}_{i,m}, \eta^{z,1}_{j,i,u}, \eta^{z,2}_{j,i,u}, \eta^{z}_{u}, \eta^{z}_{i}\}$ and the penalized parameters $\bm{\rho}=\{\rho_\alpha, \rho_z\}$ are updated by the following expressions:
        \begin{alignat*} {2}
            &\eta^{\alpha,1}_{i,m} \leftarrow \eta^{\alpha,1}_{i,m} + \frac{1}{\rho_\alpha} \cdot \alpha_{i,m} (\Tilde{\alpha}_{i,m}-1) , ~\forall i,m,\\
            &\eta^{\alpha,2}_{i,m} \leftarrow \eta^{\alpha,2}_{i,m} + \frac{1}{\rho_\alpha} \cdot (\alpha_{i,m}  -\Tilde{\alpha}_{i,m} ), ~\forall i,m,\\
            &\eta^{z,1}_{j,i,u} \leftarrow \eta^{z,1}_{j,i,u} + \frac{1}{\rho_z} \cdot z_{j,i}^u \big(\Tilde{z}_{j,i}^u-1\big), ~\forall j,i,u,\\
            &\eta^{z,2}_{j,i,u} \leftarrow \eta^{z,2}_{j,i,u} + \frac{1}{\rho_z} \cdot \big(z_{j,i}^u  -\Tilde{z}_{j,i}^u\big), ~\forall j,i,u,\\
            &\eta^{z}_{u} \leftarrow \eta^{z}_{u} + \frac{1}{\rho_z} \cdot \bigg(\sum_j \sum_i z_{j,i}^u -1\bigg), ~\forall u,\\
            &\eta^{z}_{i} \leftarrow \eta^{z}_{i} + \frac{1}{\rho_z} \cdot \bigg(\sum_j \sum_u z_{j,i}^u -1\bigg), ~\forall i,\\
        &\rho_\alpha \leftarrow c \rho_\alpha \text{ and } \rho_z \leftarrow c \rho_z,
        \end{alignat*}
        where $0<c<1$. 
        
        We define the indicators of constraint violation as follows:
        \begin{alignat*} {2}
        h(\bm{\alpha}, \Tilde{\bm{\alpha}}, \mathbf{z}, \mathbf{\Tilde{z}})=\max\bigg[\{&|\alpha_{i,m} (\Tilde{\alpha}_{i,m}-1)|, |\alpha_{i,m}  -\Tilde{\alpha}_{i,m}|, \forall i,m\}, 
        \\
        &\{|z_{j,i}^u \big(\Tilde{z}_{j,i}^u-1\big)|, |z_{j,i}^u  -\Tilde{z}_{j,i}^u|, \forall j,i,u\},  
         \\
        \big\{ \big|\sum_j \sum_i z_{j,i}^u &-1\big|, \forall u \big\}, \big\{ \big|\sum_j \sum_u z_{j,i}^u -1\big|, \forall i\big\} \bigg].
        \end{alignat*}
        By comparing $h(\bm{\alpha}, \Tilde{\bm{\alpha}}, \mathbf{z}, \mathbf{\Tilde{z}})$ with the predefined tolerance for accuracy, we can ascertain when to terminate the outer loop.

	\subsection{Overall Algorithm Description and Analysis} 
	 Building on the groundwork laid out in previous sections, we present the proposed overall PDD-based algorithm. In each iteration of the inner loop, subproblems (\textbf{SP1}), (\textbf{SP2}), (\textbf{SP3}), and (\textbf{SP4.1}) are iteratively updated with the other variables held constant. Subsequently, the dual variables and penalized parameters are updated in the outer loop. Note that we denote the objective function in (\textbf{P2}) by $\mathcal{E}$ and that of the $n$-th iterations by $\mathcal{E}^{(n)}$ (i.e., $\mathcal{E} = \frac{1}{I} \sum_i t_{\mathsf{serv},i}$ and $\mathcal{E}^{(n)} = \mathcal{E} (\bm{\alpha}^{(n)}, \mathbf{\Tilde{\bm{\alpha}}}^{(n)}, \mathbf{P}^{(n)}, \mathbf{q}^{(n)}, \mathbf{z}^{(n)}, \mathbf{\Tilde{z}}^{(n)})$). In the outer loop, $\bm{\eta}$ and $\bm{\rho}$ are updated. The details are summarized in Algorithm \ref{alg:overall}.  
    As outlined in  the analysis found in \cite{ref:PDD}, the double-loop PDD approach may find the KKT (stationary) point under Robinson's condition. However, since the original problem (\textbf{P1}) is a MINLP problem, proving compliance with Robinson's condition is challenging. Consequently, we numerically demonstrate the convergence and the optimality gap of Algorithm \ref{alg:overall} in Section \ref{sec:simulation}.
    
	\begin{algorithm}[t]
		\caption{Overall Proposed PDD-based Algorithm }
		\begin{algorithmic}
			\STATE Initialize $\{ \bm{\alpha}^{(0)}, \bm{\Tilde{\alpha}}^{(0)}, \mathbf{P}^{(0)}, \mathbf{q}^{(0)}, \mathbf{z}^{(0)}, \mathbf{\Tilde{z}}^{(0)}\}$, $\mathcal{E}^{(0)}$. Set the outer loop criterion $\epsilon_1$, the inner loop criterion $\epsilon_2$.
                \STATE Initialize outer loop count $m = 1$.
			\REPEAT 
                \STATE Initialize inner loop count $n = 1$.
			\REPEAT 
			\STATE  Update $\bm{\alpha}^{(n)}, \bm{\Tilde{\alpha}}^{(n)}$ by solving (\textbf{SP1}).
			\STATE  Update $\mathbf{P}^{(n)}$ by solving (\textbf{SP2}).
			\STATE  Update $\mathbf{q}^{(n)}$ by solving (\textbf{SP3}).
			\STATE  Update $\mathbf{z}^{(n)}, \mathbf{\Tilde{z}}^{(n)}$ by solving (\textbf{SP4.1}).
			\STATE Calculate $\mathcal{E}^{(n)} =\mathcal{E} (\bm{\alpha}^{(n)}, \mathbf{\Tilde{\bm{\alpha}}}^{(n)}, \mathbf{P}^{(n)}, \mathbf{q}^{(n)}, \mathbf{z}^{(n)}, \mathbf{\Tilde{z}}^{(n)})$
			\STATE Update inner loop count  $n = n + 1$.
    			\UNTIL{$|\mathcal{E}^{(n)} - \mathcal{E}^{(n-1)}| \leq \epsilon_2$ or $n > n_\mathsf{max}$.}
                \STATE  Update $\bm{\eta}$ and $\bm{\rho}$.
			\STATE Update outer loop count  $m = m + 1$.
			\UNTIL{$h(\bm{\alpha}, \Tilde{\bm{\alpha}}, \mathbf{z}, \mathbf{\Tilde{z}}) \leq \epsilon_1$ or $m > m_\mathsf{max}$.}
		\end{algorithmic}
		\label{alg:overall}
	\end{algorithm}

The computation complexity of the proposed algorithm depends on the four subproblems at each iteration.  
    To solve the convex optimization problems (\textbf{SP1}), (\textbf{SP2}), (\textbf{SP3}) and (\textbf{SP4.1}), interior point method is employed by CVX software \cite{ref:Kang}, whose complexities are $\mathcal{O}((IM)^{3.5} \log(1/\epsilon))$, $\mathcal{O}((IM)^{3.5} \log(1/\epsilon))$, $\mathcal{O}((2M)^{3.5} \log(1/\epsilon))$ and $\mathcal{O}((IJU)^{3.5} \log(1/\epsilon))$, respectively.
    Therefore, the overall complexity can be expressed as $\mathcal{O} (m_\mathsf{max} n_\mathsf{max} ( (IM)^{3.5} \log(1/\epsilon)+ (IJU)^{3.5} \log(1/\epsilon) ))$.

    Notably, the search space for binary variables escalates exponentially as $\mathcal{O}(2^{I^2 JMU})$. Adopting a polynomial-complexity algorithm makes the network operation feasible. 
    In addition, our framework targets IoT environments with minimal dynamicity to maximize long-term performance, enabling the identification of sustainable solutions that once established, alleviate practical implementation concerns. This efficiency underscores our approach's ability to manage computational complexity while ensuring operational feasibility.

\begin{figure}[t]
		\centering
		{\includegraphics[width=0.7\columnwidth]{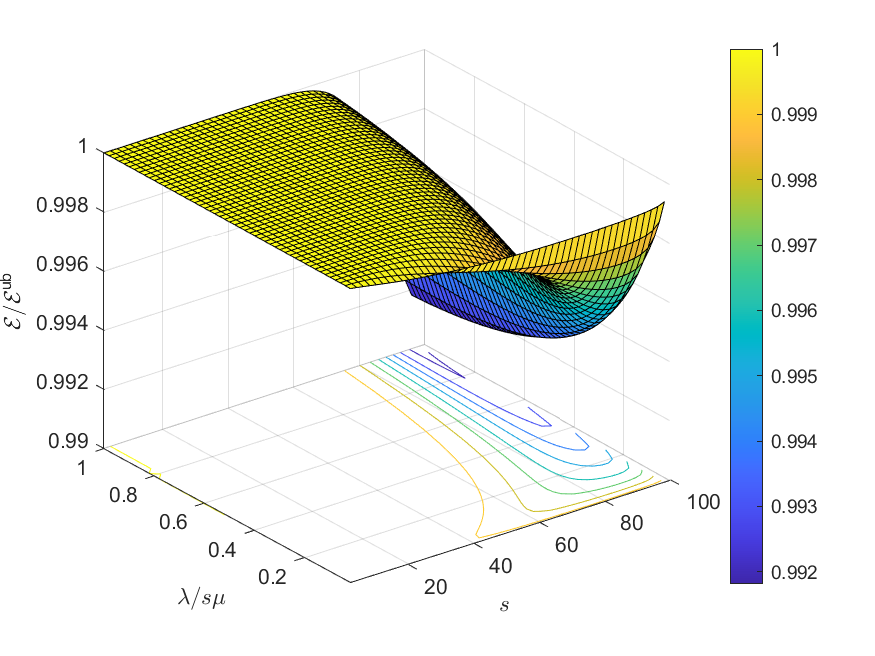}} 
		\caption{Ratio between the expected user service delay and its upper bound.} 
		\vspace{-10pt}
		\label{fig:ratio}
	\end{figure}

	\begin{remark} \label{rem:conv}
        Fig. \ref{fig:ratio} depicts the ratio $\frac{\mathcal{E}}{\mathcal{E}^\mathsf{ub}}$ against the number of computing units $s$ and the traffic intensity $\frac{\lambda}{s\mu}$, with a fixed total service rate $s\mu$ of 100. This demonstrates that the user service delay $\mathcal{E}$ closely approximates its upper bound $\mathcal{E}^\mathsf{ub}$, with a ratio exceeding 0.99. Notably, as $s \rightarrow 2$, $\mathcal{E}$ equals $\mathcal{E}^\mathsf{ub}$, indicating an extremely tight upper bound. For convergence, we limit the number of computing units $s$ to small values, e.g., $s=2$. This is feasible by splitting or merging computing units into a small number of virtual units with the same computing power, maintaining the total service rate $s\mu$ \cite{ref:Cheng}.
	\end{remark}
	
	\section{Numerical Results and Discussions} \label{sec:simulation}
	
	\begin{table}[t]
	\caption{Network system parameters
		\label{tab:env}}
	\centering
	\begin{tabular}{|c|c|}
		\hline
		\textbf{Parameters} & \textbf{Settings} \\
		\hline \hline
		Number of IoTs, $I$ & $20$\\
		\hline
		Number of MECs, $J$ &  $4$\\
		\hline
		Number of UAVs, $M$ &  $3$\\
		\hline
		Number of computing units, $s$ & $2$\\
		\hline
		Service rate of each computing unit, $\mu$  & $4$ tasks/sec\\
		\hline
		Average user request rate, $\lambda_\text{avg}$ & $1.2$ tasks/sec \\
		\hline
		\hline
		\hline
		Bandwidth, $B$ & $1$ GHz\\
		\hline
		Noise spectral density, $N_0$ & $-174$ dBm \\
		\hline
		Task input data size, $D_\mathsf{in}$ & $10$ MB \\
		\hline
		Transmit power (IoT), $P_\mathsf{IoT}$ & $200$ mW \\
		\hline
		Transmit power (UAV), $P_\mathsf{UAV}$ & $2$ W \\
		\hline
		UAV altitude, $H$ & $20$ m \\
		\hline
		\hline
		Heights of MEC and IoT, $h_\mathsf{MEC}$, $h_\mathsf{IoT}$ & $3.0$ m, $0.3$ m\\
		\hline
		Height and radius of blockers, $h_\mathsf{b}$, $\tau_\mathsf{b}$ & $1.7$ m, $0.3$ m \\
		\hline
		Density of blockers, $\beta_\mathsf{b}$ & $0.2 \text{ m}^{-2}$ \\
		\hline
	\end{tabular}
\end{table}

   	In our network configuration, we consider 20 IoTs, 4 MEC servers and 3 UAVs within $400$m $\times$ $400$m squared area network, where IoTs and MECs are uniformly distributed and the locations of UAVs are initialized using the $K$ mean clustering algorithm.
    The UAVs are assumed to be located at a fixed altitude of $20$m. We select the reference carrier frequency $f_o$ at $0.34$ THz to evaluate, which is known that there is a large transmission window with no path loss peak \cite{ref:Wang2}, and the bandwidth of $1$GHz for each sub-band $u \in \mathcal{U}$. 
    In addition, the molecular absorption coefficient $K(f)$  is referred to the HITRAN database \cite{ref:HITRAN}. The specific environmental details of our simulations are given in Table \ref{tab:env}, unless otherwise stated. Besides, network topology which is statistically distributed are averaged after 50 simulation runs. To evaluate our proposed algorithm, we adopt the following baseline algorithms:
	\begin{enumerate}
	    \item \textit{UAV optimization (\textbf{UO}):} Optimizing UAV-related design variables $\bm{\alpha}, \mathbf{P}$, and $\mathbf{q}$ by solving subproblems (\textbf{SP1}), (\textbf{SP2}), and (\textbf{SP3}) iteratively. The IoT's tasks are offloaded to the nearest MEC server without exceeding the total service rate. 
	    \item  \textit{User-resource associations optimization (\textbf{UAO}):} Only optimizing the user-resource associations $\mathbf{z}$ by resolving subproblem (\textbf{SP4.1}) with UAV selection when the IoT-MEC distance surpasses that of IoT-UAV, and using equal power allocation.
	    
        \item \textit{No relay with successive convex approximation based user-resource association optimization (\textbf{NR-SCA}):}         No UAV relay is considered and the user-resource association optimization is conducted using the successive convex approximation (SCA), as described in \cite{ref:Alvi}. 
        
	    \item \textit{UAV optimization and genetic algorithm based user-resource association optimization (\textbf{UO-GUAO}):} The genetic algorithm based user-resource associations \cite{ref:Song} is adopted with the proposed UAV optimization, where iteratively solving subproblems (\textbf{SP1}), (\textbf{SP2}), (\textbf{SP3}).
     
        \item \textit{Block coordinate descent and successive convex approximation based optimization (\textbf{BCD-SCA}):} A block coordinate descent based algorithm \cite{ref:Kang,ref:Diao} is adopted, where subproblems (\textbf{SP1}), (\textbf{SP2}), (\textbf{SP3}), and (\textbf{SP4.1}) are iteratively solved without double-loop structure. The solutions to (\textbf{SP1}) and (\textbf{SP2}) are obtained using \textbf{Theorems \ref{thm:sp1}} and \textbf{\ref{thm:sp2}}, respectively. Additionally, the SCA based user-resource association is conducted \cite{ref:Alvi}.

	\end{enumerate}
 
    In addition, in order to analyze the optimality gap, we adopt a method of discretizing the continuous variables by quantizing them with $n_{q1}$ and $n_{q2}$ for a network of  size $l_\text{net} \times l_\text{net}$, where $P_{i,m} \in \{0, \Delta_1, 2\Delta_1, ..., P_\text{UAV}\}$ and $x_m, y_m \in \{0, \Delta_2, , 2\Delta_2, ..., l_\text{net}\}$, i.e., $\Delta_1 = P_\text{UAV} / (n_{q1}-1)$ and $\Delta_2 = l_\text{net} / (n_{q2}-1)$. 
        In our simulation, we set the quantization levels as $n_{q1}=20$ and $n_{q2}=40$, resulting in a total of 1,600 quantized points in the UAV deployment.  
       
	\begin{figure}[t]
		\centering
		\subfigure [] 
		{\label{fig:conv}\includegraphics[width=0.7\columnwidth]{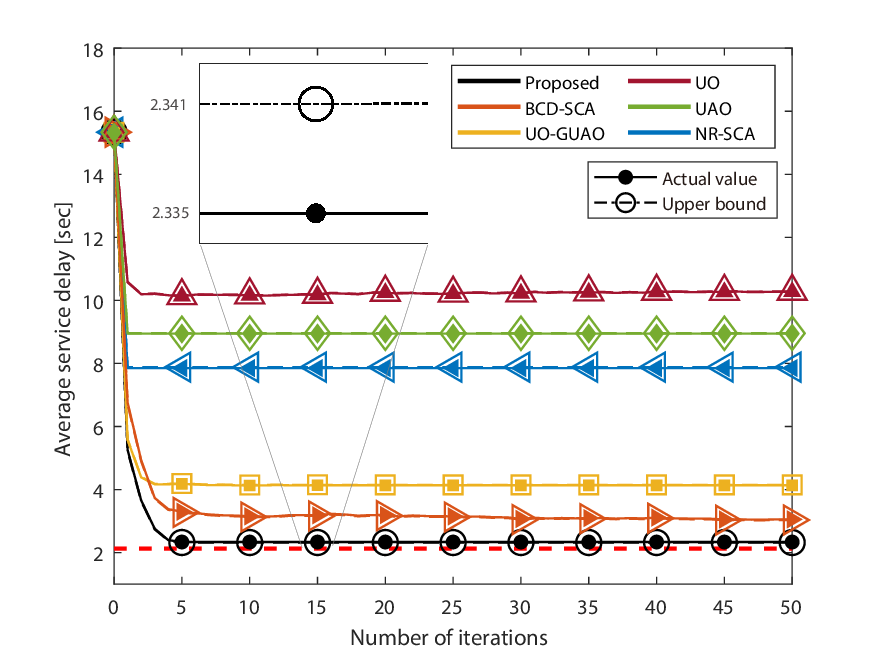}} 
		\hfil
		
		\subfigure [] 
		{\label{fig:delay}\includegraphics[width=0.7\columnwidth]{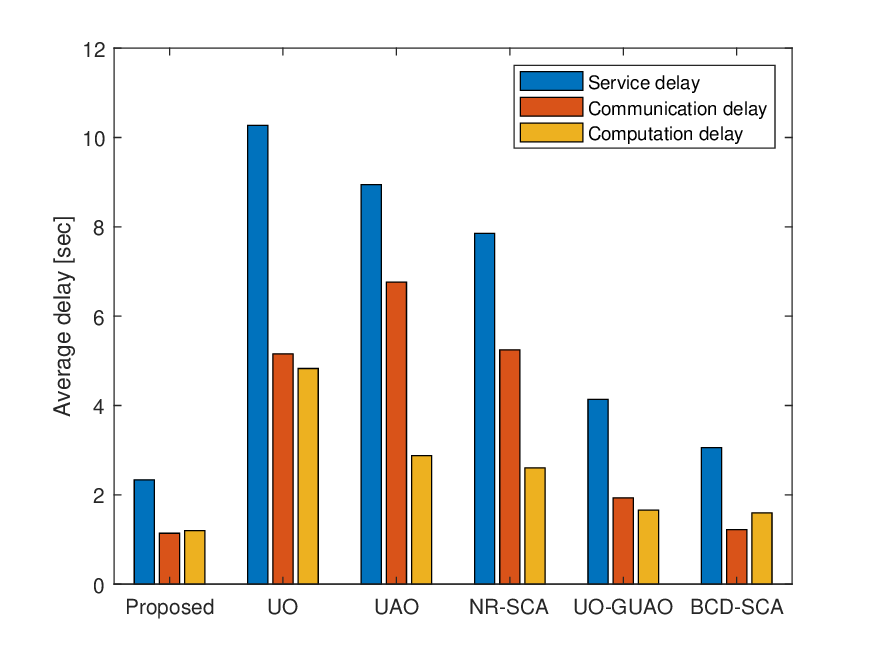}}
		
		\caption{(a) Convergence of the proposed algorithm and its upper bound, (b) communication and computation delay of the converged user service delay.}
		\vspace{-10pt}
		\label{fig:convres}
	\end{figure}
 
    In Fig. \ref{fig:conv}, we present the convergence of our proposed design and compare it with benchmarks. It is shown that the PDD-based proposed algorithm and benchmarks reach the local optimum within a finite number of iterations. However, our proposed solution outperforms the others. Furthermore, the upper-bounded solutions, denoted by the dash-dotted line in the figure, are very close to their exact values. Specifically, the proposed solution converges to 2.3354 seconds, and its upper bound is 2.3509 seconds, resulting in a ratio of $\frac{\mathcal{E}}{\mathcal{E}^\mathsf{ub}} \thickapprox 0.9934$. Therefore, the solutions can be considered approximately equivalent. Comparatively, the UAO optimizes $t_\mathsf{comp}$, while the UO primarily enhances $t_\mathsf{comm}$. Consequently, in a general context, the UAO tends to outperform the UO, given the exponential increase in queueing delay when MEC becomes overloaded. 
    NR-SCA outperforms UO and UAO, revealing that, in scenarios where UO and UAO are not jointly optimized, it is more advantageous to transmit without the use of UAV relays. Conversely, alternative methods, such as BCD-SCA and UO-GUAO, demonstrate superiority over NR-SCA, indicating that UAV relays can prevent overloads while reducing communication delays, thereby enhancing performance. Although the genetic algorithm-based UAO and BCD-SCA converge to a suboptimal value that is lower than those achieved by UO, UAO, and NR-SCA, our proposed solution achieves a more favorable suboptimal outcome. Moreover, the red dotted line in the figure represents the optimal solution driven by the exhaustive search mentioned earlier, with a value of 2.1253 seconds. As in Fig. \ref{fig:conv}, our proposed technique finds solutions that are close to optimal with polynomial time complexity for system parameters.

	More specifically, Fig. \ref{fig:delay} derives the communication and computation delay of converged user service delay for each scheme. As aforesaid, UO alleviates the communication delay by adjusting the UAV selection, power allocation, and the placement, but cannot handle the task overloading. Conversely, the UAO efficiently manages task overload, minimizes computation delay, but doesn't fully harness the benefits of optimizing UAVs for communication delay.
    Meanwhile, the proposed algorithm, the BCD-SCA, and the UO-GUAO optimize both jointly, obtaining the performance improvement. Notice that the genetic-based UAO can approach to the improved suboptimum with the longer generations or populations, that might require the massive computational complexity.
	
	\begin{figure}[t]
		\centering
		\subfigure [] 
		{\label{fig:top}\includegraphics[width=0.7\columnwidth]{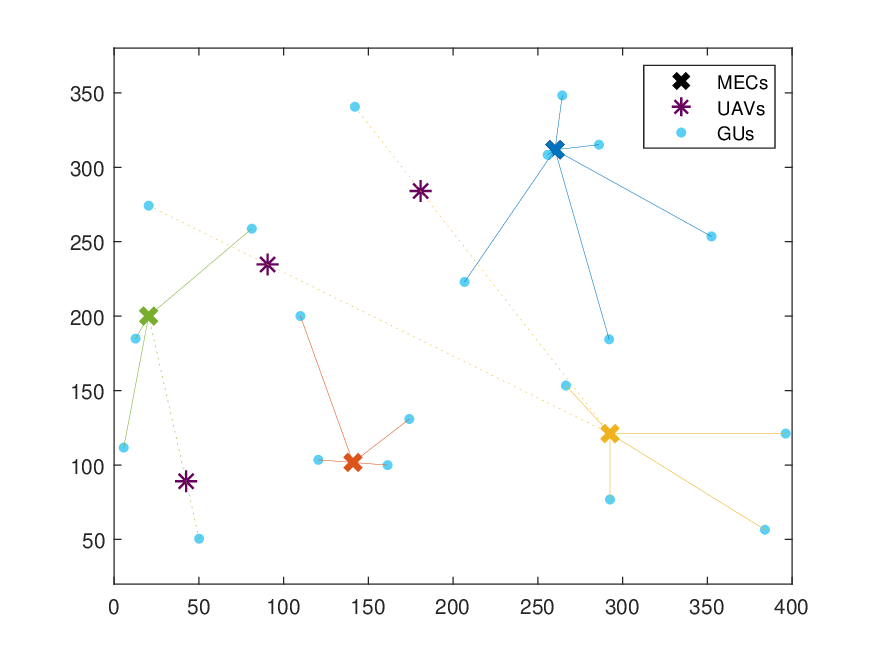}} 
		\hfil
		
		\subfigure [] 
		{\label{fig:servuse}\includegraphics[width=0.7\columnwidth]{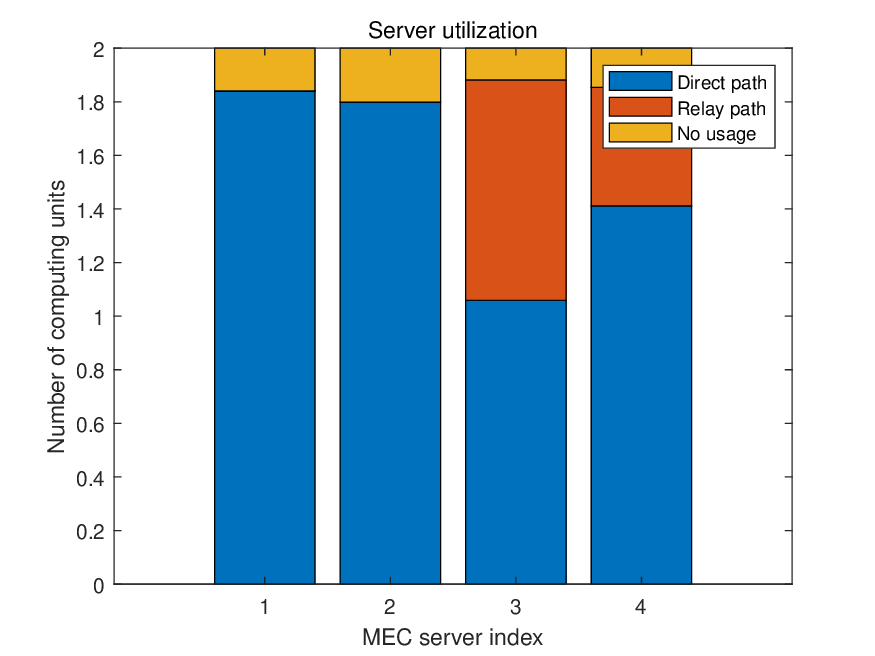}}
		
		\caption{(a) An example of network topology and (b) server utilization.}
		\vspace{-10pt}
		\label{fig:result}
	\end{figure}
	 Fig. \ref{fig:result} describes an example of (a) resulted UAV positioning and IoT's transmission links, (b) the server utilization, for a given network topology. Colored x marks, purple asterisk marks, and light blue dots describe the MECs, UAVs, and IoTs each. Also, line and dashed line represent direct link and relay link, respectively. As depicted in Fig. \ref{fig:top}, two IoTs positioned at the upper-left corner offload to the distant MEC with index 3 (yellow x mark) using a UAV relay. To minimize transmission delay, the logical choice would be to associate with the nearest MEC, such as MEC 1 (blue x mark) or 4 (green x mark). However, due to overloading in these MECs, queueing delay might drastically rise. To address this, the IoTs opt to offload tasks to more distant yet feasible MECs, like MEC 3, for load balancing. Similarly, the IoT situated in the lower-left area employs a UAV relay to counter significant path loss. Fig. \ref{fig:servuse} presents MEC server utilization, distinguishing between direct link offloading (blue) and relay path (orange) for server usage. The server queue becomes unstable when task offloading surpasses computing capacity, indicated by the absence of a yellow bar in Fig. \ref{fig:servuse}. In this context, Fig. \ref{fig:servuse} illustrates that every MEC server maintains a well-balanced load, resulting in a stabilized computation delay.

	\begin{figure}[t]
		\centering
		{\includegraphics[width=0.7\columnwidth]{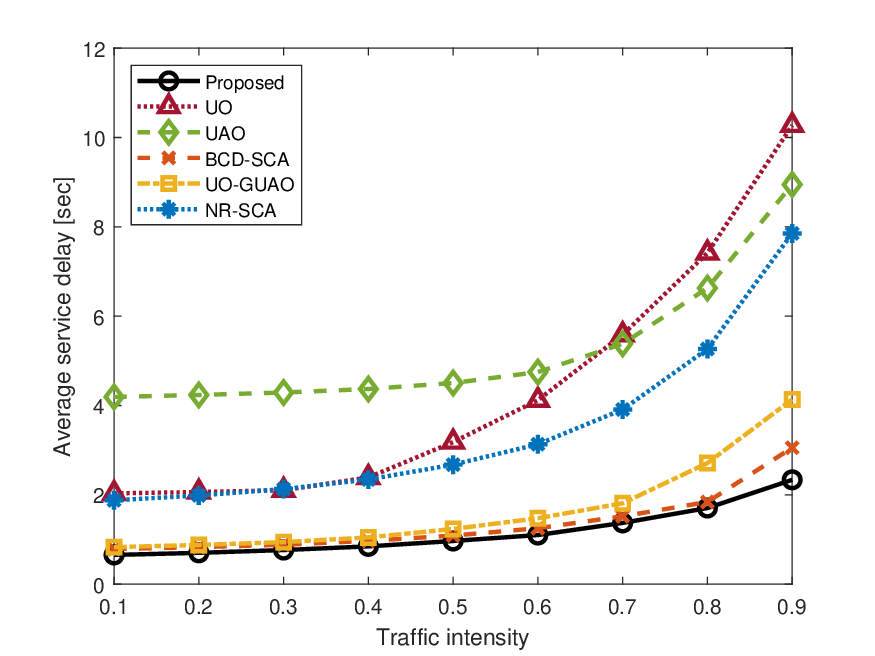}} 
		\caption{Expected user service delay with respect to traffic intensity.}
		\vspace{-10pt}
		\label{fig:traffint}
	\end{figure}
	Fig. \ref{fig:traffint} offers the performance comparison with respect to the traffic intensity $\frac{\lambda}{s\mu}$, measuring the intensity of the total offloaded tasks by comparison with the server capacity. As the traffic intensity grows, becoming computation-limited, the expected user service delay for all schemes degrades owing to the increment of queue waiting delay. Notably, at the low intensity region, the delay of both UO is lower than that of UAO due to the small queueing delay. But UAV optimized schemes are defeated by UAO in the computation-limited regime, since the UO cannot manage the overloaded tasks, which leads to the poor queueing delay. On the contrary, the UAO controls the IoT-MEC server associations to let the traffic load balance. Thereby, it can prevent diverse queueing delays, even though the intensity approaches almost one. But, the UAO still suffers from the communication delay. The proper UAV selection, power control, and placement can further improve the user service delay. Since the relay management and traffic balancing are carried out in the proposed one, the BCD-SCA, and the UO-GUAO, the performance enhancement is achieved. Still, the proposed method attains a better quality solution than the BCD-SCA and the UO-GUAO.

	\begin{figure}[t]
		\centering
		{\includegraphics[width=0.7\columnwidth]{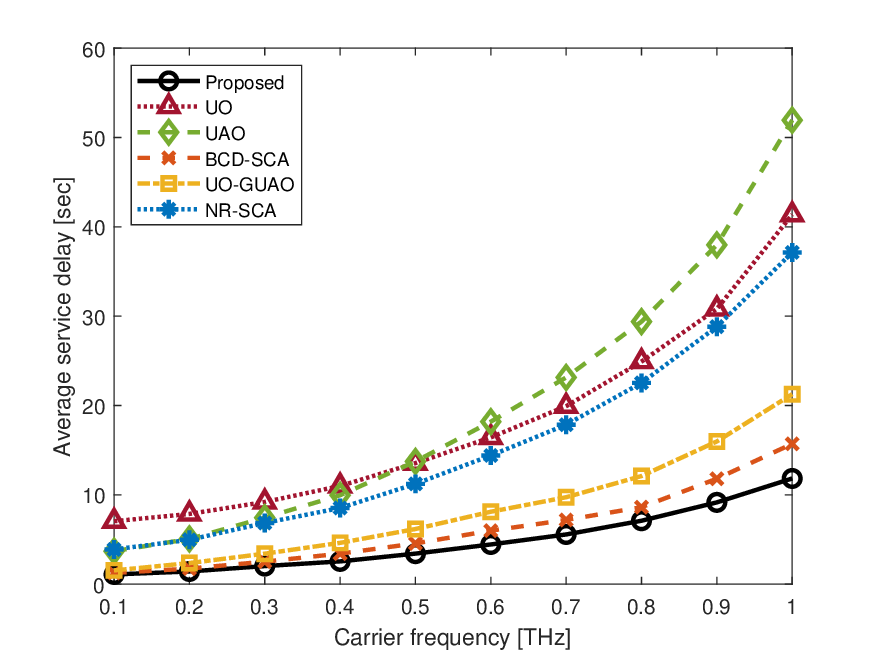}} 
		\caption{Expected user service delay with respect to the carrier frequency.} 
		\vspace{-10pt}
		\label{fig:freq}
	\end{figure}
	Fig. \ref{fig:freq} compares the converged expected user service delay with different baselines with respect to the reference carrier frequency $f_o$ from 0.1THz to 1THz. In high frequency band, it is vulnerable to the molecular absorption loss, hence all the methods are degraded. Especially, the UO outperforms the UAO in the communication-limited regime, where the path loss becomes severe. This is because the IoTs become more dependent on the UAV relay due to the severe path loss, but the UAO is not able to control the transmission paths of all associations by the UAV optimization. On the flip side, the computation delay is dominant in low frequency band, the UAO takes lower delay. Lastly, we can see that the proposed one outperforms the others.
	
	\begin{figure}[t]
		\centering
		{\includegraphics[width=0.7\columnwidth]{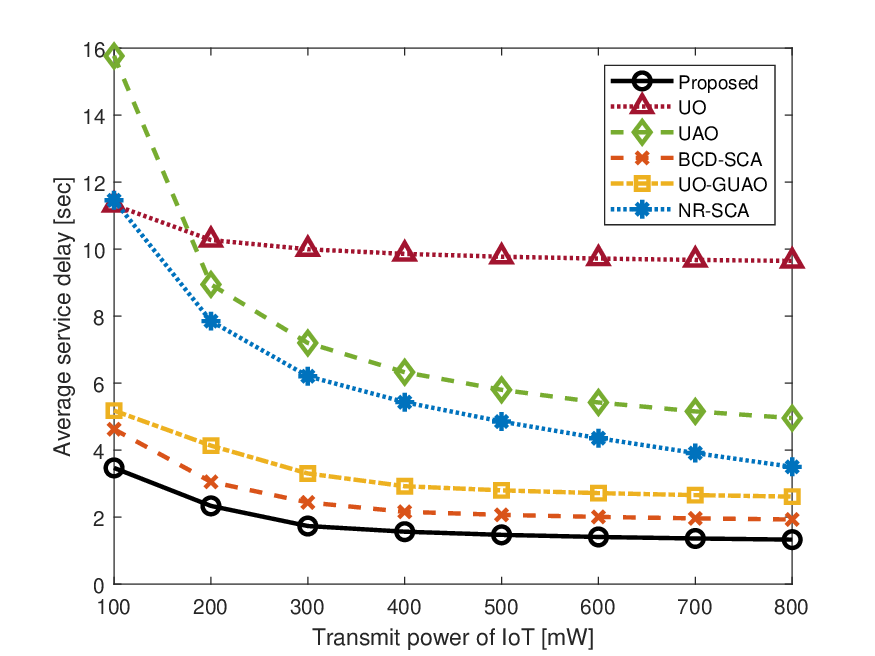}} 
		\caption{Expected user service delay with respect to the transmit power of IoT.} 
		\vspace{-10pt}
		\label{fig:piot}
	\end{figure}
	Fig \ref{fig:piot} illustrates the expected user service delay with respect to the transmit power of IoTs, i.e., $P_\mathsf{IoT}$. Notably, UAO exhibits a significant advantage in this context. Intuitively, at lower transmit power levels, most IoTs tend to rely on the UAV relay, making UAV optimization crucial. Conversely, at higher transmit power levels, direct communication becomes more advantageous, reducing the reliance on UAV relays. Consequently, NR-SCA performs well in high transmit power scenarios, approaching the performance of UO-GUAO. The proposed solution, along with BCD-SCA and UO-GUAO, accounts for both factors, resulting in a lower expected user service delay compared to the other three schemes. 
	
	From all these aspects, we emphasize that both communication and computation optimizations are vital in the MEC network, no matter what environmental parameters are limited. Our proposed design provides a well-optimized solution to minimize the overall expected user service delay, not just to minimize either communication or computation delay solely.

	\section{Conclusion} \label{sec:conclusion}
	In this article, we proposed a novel architecture of THz-enabled MEC system with multi-UAV communication relays. To minimize user service delay therein, a joint optimization problem of UAV selection, power allocation, deployment, as well as the user offloading-and-sub-band associations was formulated. To deal with the formulated MINLP problem, it is decomposed into four subproblems and solved via iterative PDD algorithm. The simulation results showed that the proposed algorithm can find a high-quality suboptimum in polynomial time order, and outperform the benchmarks.

\end{document}